\documentclass[10pt,
 reprint,
 onecolumn,
superscriptaddress,
nofootinbib,
 amsmath,amssymb,
 aps,
prd,
]{revtex4-2}
\usepackage{hyperref}
\usepackage[normalem]{ulem}
\usepackage{color}
\usepackage{amsthm}
\usepackage{graphicx}
\usepackage{dcolumn}
\usepackage{bm}
\usepackage{mathrsfs}
\usepackage{dsfont}
\usepackage{selinput}
\usepackage{mathtools}
\def\d{{\rm d}}
\usepackage{dcolumn}
\usepackage{setspace}
\usepackage{xcolor}

\newcommand{\lk}{l(l+1)}
\newcommand{\hor}{\mathcal{H}}
\newcommand{\xzero}{\lambdabar}

\begin{document}

\title{
Radiative properties of a nonsingular black hole: Hawking radiation and gray-body factor
}

\author{Asier Alonso-Bardaji}
 \email{asier.alonso@ehu.eus}
\affiliation{
Department of Applied Mathematics and EHU Quantum Center, University of the Basque Country UPV/EHU,\\
Plaza Ingeniero Torres Quevedo 1, 48013 Bilbao, Spain
}%
 
\author{David Brizuela}
 \email{david.brizuela@ehu.eus}

\author{Marc Schneider}
\email{marc.schneider@ehu.eus}
\affiliation{
Department of Physics and EHU Quantum Center, University of the Basque Country UPV/EHU,\\
Barrio Sarriena s/n, Leioa 48940, Spain}

\begin{abstract}
We study the radiative properties of a spherical and singularity-free black-hole geometry recently proposed in the literature. Contrary to the Schwarzschild spacetime,
this geometry is geodesically complete and regular, and, instead of the singularity, it presents
a minimal surface that connects a trapped (black-hole) with an antitrapped (white-hole) region.
The geometry is characterized
by two parameters: the Schwarzschild radius and another parameter that measures the area of
the minimal surface. This parameter is related to certain corrections expected in the
context of loop quantum gravity to the classical general-relativistic dynamics. We explicitly
compute the spectrum of the Hawking radiation and the gray-body factor.
Since the gravitational potential is shallower than in Schwarzschild,  
the emission spectrum turns out to be colder and purer (less gray).
From this, we sketch the evaporation history of this geometry and conclude that,
{ under certain assumptions,} instead of completely
evaporating, the black hole naturally leads to a remnant, which provides a possible resolution to the
information-loss issue.
\end{abstract}

\maketitle

%\large

\section{Introduction}

General relativity (GR), together with quantum mechanics, stands as one of the most successful theories in modern physics. It provides a remarkably accurate description of the observable universe, modeling it as a spacetime consisting of a four-dimensional Lorentzian manifold. General relativity has been confirmed through numerous observations, but there are strong indications pointing towards its incompleteness. 
The singularity theorems \cite{Penrose:1964wq,Hawking:1970zqf,Senovilla:2021pdg} show that Einstein’s equations fail to describe the most extreme regions of the cosmos: the beginning of the universe and the deep interior of black holes. 

It is generally believed that a quantum theory of gravity will eventually resolve these issues. Today,
loop quantum gravity is one of the candidates for a quantum theory of gravity \cite{Rovelli:2004tv,Ashtekar:2004eh,Thiemann:2007pyv,Ashtekar_2021}. Among its predictions is the discrete spectrum of geometric quantities, such as area and volume \cite{Rovelli:1994ge,Abhay}, which has been used, for example, in calculations of black-hole entropy \cite{Meissner:2004ju,Rovelli:1996dv,Agulló_2012,BarberoG:2022ixy,Perez:2017cmj}, and in resolving the Big-Bang singularity in cosmological models \cite{Ashtekar:2003,Bojowald:2008}. However, the theory remains incomplete, with the formulation of its full dynamics still being an open problem. This has motivated significant efforts to develop effective models, aiming to capture the main quantum features of gravity and to shed light on the most elusive aspects of the theory. 

Nonetheless, the absence of direct experimental evidence is a major obstacle to build and test effective theories.
Fortunately, in cosmological models, effective theories have proven successful in predicting expected effects of
the full theory of loop quantum gravity \cite{lqcreport,lqcbrief,Rovelli_2014}. Generalizing such predictions to less symmetric scenarios would endorse the idea that classical singularities are resolved. In particular, much work has been performed in spherical symmetry \cite{Kelly:2020uwj,Ashtekar:2018cay,Ashtekar:2018lag,Bodendorfer:2019cyv,Bodendorfer:2019nvy,BenAchour:2018khr,Gambini:2020nsf,Gambini:2020qhx,Alonso-Bardaji:2021tvy,Alonso-Bardaji:2021yls,Alonso-Bardaji:2022ear,Alonso-Bardaji:2023niu,Gambini:2022dec,Ben_tez_2020,Giesel_2021,Gambini_2022,Husain_2022,Giesel_2023,Cipriani_2024,Alonso-Bardaji:2023qgu,Alonso-Bardaji:2024tvp}, but a key challenge is the reconciliation between the discrete spacetime structure predicted by loop quantum gravity and the continuous picture implied by diffeomorphism symmetry in general relativity \cite{Alonso-Bardaji:2023vtl,Bojowald:2023xat}.
This is the main framework in which the present research is conducted.

In this paper
we study the radiative features of a recent generalization of the Schwarzschild black hole \cite{Alonso-Bardaji:2021yls,Alonso-Bardaji:2022ear} for which the singularity is replaced by a transition surface that
leads into a time-reversed region. In other words, the black-hole region evolves into a white hole that emerges into a parallel universe. Apart from the constant radius of the horizon $r_g$, the spacetime
is characterized by an additional parameter $r_0$ that measures the area of the transition surface.
It is important to point out that
this geometry represents no ad hoc construction; instead, it is a solution of the equations of motion generated by a Hamiltonian
constraint that is deformed with respect to the GR Hamiltonian \cite{Alonso-Bardaji:2021yls,Alonso-Bardaji:2022ear}. The deformation
functions are trigonometric functions, and thus the additional parameter can be understood
as including holonomy corrections in the model, which are expected from the
regularization performed
in the context of loop quantum gravity. 
This is why we will occasionally refer to the corrections of the model with respect to GR
as quantum-gravity corrections.
However, the deformed model is more
general and it is not directly derived from the theory of loop quantum gravity.

In any case, independent of their fundamental origin,
the presence of corrections with respect to GR motivates the natural question:
Are these (quantum-gravity) modifications measurable by an outside observer?
A natural pathway to answer this question is through quantum field theory in curved spacetimes, as this theory has a long
and fruitful history of probing spacetime dynamics in the presence of test quantum fields.
In particular, in 1974, Hawking predicted that black holes create particles at the horizon \cite{hawking1974,hawking1975}. The emission rate follows Planck's law, and, thus, its temperature scales anti-proportional with the size of the black hole. However, this phenomenon includes deep conceptual challenges, e.g., the information-loss issue.
The Hawking radiation implies a net energy loss of the black hole, which may lead to its complete evaporation.
The classical singularity would be naked and it
is not clear what would happen with all the information previously contained within the horizon.

However, quantum-gravity corrections are expected to become significant during the final stages of evaporation,
when the black hole's size approaches the Planck scale. These corrections may drastically alter the final stages of evaporation
and directly impact on the fate of the information. In particular,
some scenarios propose that, instead of completely evaporating,
black holes may leave behind Planck-scale sized remnants \cite{Rovelli:2024sjl}.
These compact stable objects would then retain part of the entropy and avoid classical divergences,
offering a potential resolution to the information-loss issue. Nonetheless, their dynamics and
precise properties (as well as the mechanisms that lead to their formation) remain active areas of research. 
{
In this paper we will show that, assuming that the parameter $r_0$ is constant
during the evaporation,
the model under analysis naturally leads to a remnant
of vanishing temperature and entropy.
Such assumption suggests that $r_0$ is a fundamental scale,
which is possibly associated to quantum gravity.
For instance, in Ref.~\cite{Borges:2023fub},
it was argued that $r_0$ should be related to the area gap of loop quantum gravity,
and thus Planck-size remnants occur as asymptotic configuration
of the present model.
}

While Hawking radiation is thermal for isolated black holes, the surrounding gravitational potential modifies the spectrum. Particles within the gravitational well are subjected to scattering processes that
make some of the particles emitted by the black hole to bounce at the potential
and fall back again through the horizon, instead of propagating to infinity.
These scattering effects cause a digression from the pure black-body radiation and are captured by the gray-body factor.

The Hawking spectrum and the gray-body factor provide in-principle measurable quantities that admit certain
sensitivity toward quantum-gravity modifications \cite{calza2025grayhawk,calza2025primordial1,calza2025primordial2}.
{The main goal of this paper is to derive their specific form
for a massless scalar field propagating on the regular black-hole
geometry presented in Refs.~\cite{Alonso-Bardaji:2021yls,Alonso-Bardaji:2022ear}.
In particular, based on the WKB approximation, we provide explicit analytic expressions that show the leading corrections for s-waves, 
as well as for modes with large angular momentum. This complements a recent numerical study of the gray-body factor in this model for small angular momentum and higher-spin particles \cite{Menezes:2025mfa}.}

The remainder of the paper is organized as follows.
After introducing our model in Sec.~\ref{sec.prelim}, we show
that the occurrence of the transition surface modulates the temperature
as well as the gray-body factor. More precisely,
the computation of the temperature is presented in Sec.~\ref{sec:tunneling},
while the general derivation of the gray-body factor is carried out in Sec.~\ref{sec:gbf}.
Furthermore, in Sec.~\ref{sec:examples} the properties of the gray-body factor are analyzed
for certain specific limits. In Sec.~\ref{sec.eva}, we put a special emphasis on the
last stages of the evaporation process to understand how thermodynamic concepts,
like the entropy, behave when the remnant phase is approached.
We conclude the main text with a summary and discussion of the main results in Sec.~\ref{sec:conclusions}. Finally, in the two appendices, App. \ref{app.others}
and App. \ref{app:rmax}, we present certain
technical features of the model.

Throughout the article, we work in the unit system with $G=c=k_B=1$.

\section{The nonsingular black-hole solution}\label{sec.prelim}

In Refs.~\cite{Alonso-Bardaji:2023vtl,Bojowald:2023xat} the most general
family of Hamiltonian constraints, that are
quadratic in first-order and linear in second-order radial derivatives of the phase-space variables,
and obey specific covariance conditions was presented.
This family is parametrized by seven free functions of
the areal radius. For any shape of these functions, any given solution
of the equations of motion unambiguously,
that is, independently of the chosen gauge, defines the geometry of the
spacetime. In fact, all the
Hamiltonians represent spherically symmetric spacetimes $(\mathcal{M},g)$
with topology $\mathcal{M}\simeq\mathbb{M}\times\mathbb{S}_2$,
$\mathbb{M}$ being a two-dimensional manifold, and they contain 
a Killing vector field $\xi$ in the sector $\mathbb{M}$.
If the signature of the spacetime is Lorentzian,
the causal character of $\xi$ defines either static regions
(for timelike $\xi$) or homogeneous regions (for spacelike $\xi$).
Such regions are usually separated by Killing horizons, where $\xi$ is null.

A particular member of such family of Hamiltonians is the Hamiltonian of spherical vacuum general relativity,
which leads to the Schwarzschild geometry, characterized by a constant parameter $r_g$
encoding the area of the horizon as $4\pi r_g^2$.
Other members of this family of Hamiltonians can thus be understood as implementing
covariant deformations of general relativity. As commented above, there remains still quite a lot of freedom in the model (seven
functions that can be arbitrarily selected). One simple choice of these functions
leads to the model first analyzed in detail in Refs.~\cite{Alonso-Bardaji:2021yls,Alonso-Bardaji:2022ear}.
Here, the functions are chosen to admit a sinusoidal form, and they 
depend on a
parameter $\lambda\in\mathbb{R}^+$, which, in the loop-quantum-gravity literature is usually named `polymerization parameter'
and it can be related to the fiducial 
length of the holonomies.
Independently of the interpretation of the correction terms, the most relevant part for our purpose
is that the solutions of the dynamics generated by such Hamiltonian
leads to a ``nonsingular black-hole spacetime'',
in the sense that it is qualitatively similar to a Schwarzschild black hole, though the singularity is replaced by
a minimal surface that splits the trapped region into a black-hole (future trapped) and a while-hole (past trapped) region.

The goal of this paper is to analyze the radiative properties of this particular regular black-hole solution.
The remainder of this section is divided into two subsections: In Sec.~\ref{sec.geometry} we present the metric and
briefly describe the
global geometric features of this spacetime, while in Sec.~\ref{sec.radiation} we present the basics to understand the radiative properties of the black-hole horizon.

\subsection{Geometry}\label{sec.geometry}

\begin{figure}[t]
    \centering
    \includegraphics[width=.7\linewidth]{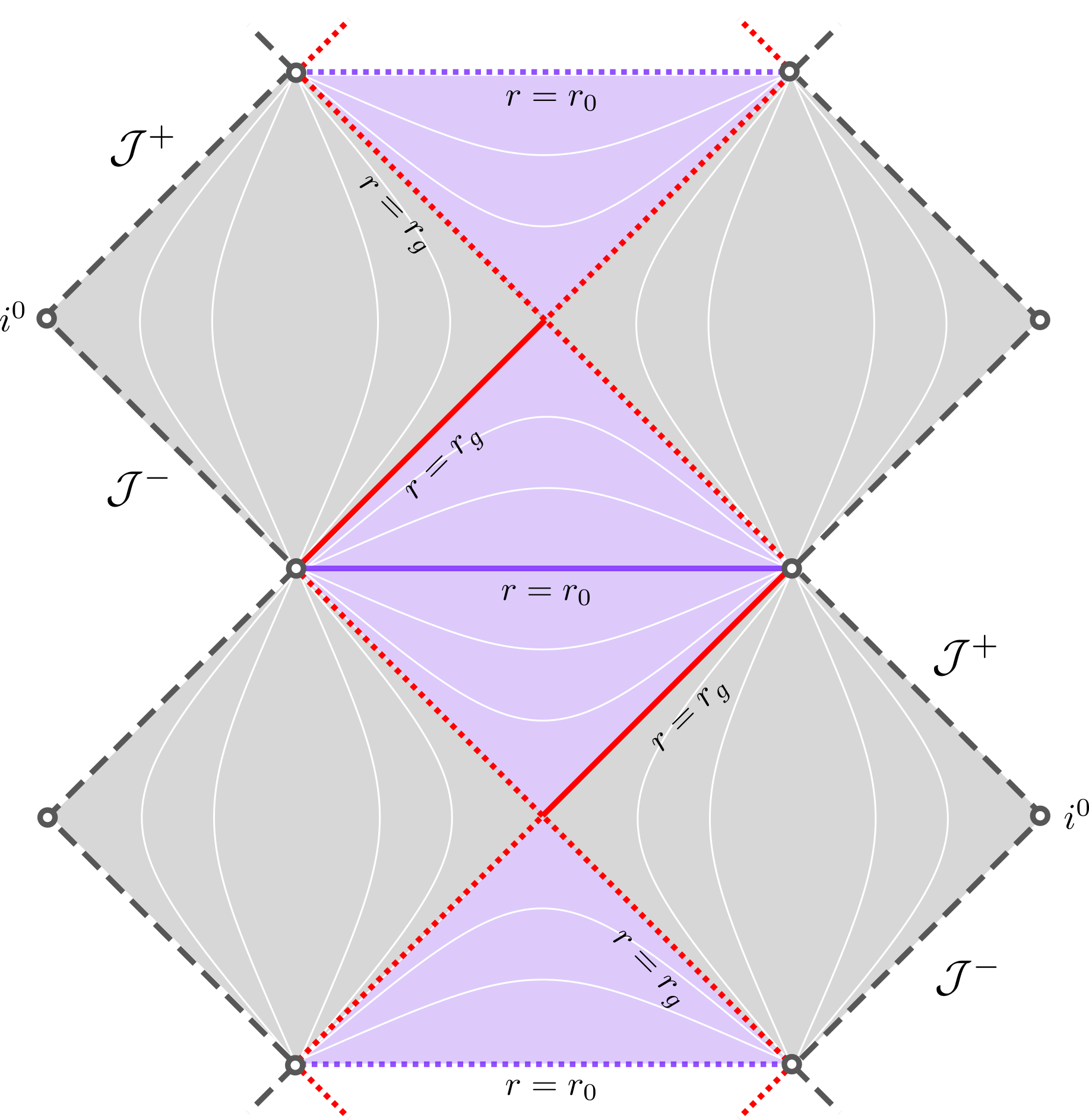}
    \caption{Maximal extension of the singularity-free geometry. The horizon $\hor$ ($r=r_g$) is depicted by the red lines at $45^\circ$. The transition surfaces $\mathcal{T}$ ($r=r_0$) are the horizontal purple lines. Some other surfaces of constant $r$ are drawn in white. These are timelike in the exterior asymptotic regions (shaded in gray), where the chart~\eqref{eq.metric} is defined, and spacelike in the interior trapped regions (shaded in purple). Dashed gray lines represent past and future null infinities, and the gray rings in their intersections are timelike and spacelike infinities.}
    \label{fig.diagram}
\end{figure}

As explained above, the vacuum solution of the commented model describes a 
globally hyperbolic, geodesically complete, and spherically symmetric
spacetime \cite{Alonso-Bardaji:2021yls,Alonso-Bardaji:2022ear}.
In a stationary gauge,
the line element that describes the exterior of this regular black hole takes the form,
\begin{align}\label{eq.metric}
    \d s^2=&\,
    -\bigg(1-\frac{r_g}{r}\bigg)\d t^2 
    +\bigg(1-\frac{r_0}{r}\bigg)^{-1}\bigg(1-\frac{r_g}{r}\bigg)^{-1}\d r^2
    +r^2(\d\vartheta^2+\sin^2(\vartheta) \d\varphi^2),
\end{align}
where $r_g$ and $r_0$ are positive constant parameters.
The former denotes the usual gravitational radius, and thus the black-hole horizon
is defined as $\hor:=\{r=r_g\}$. The latter is directly related to the polymerization parameter $\lambda\in\mathbb{R}^+$ through $r_0:=\lambdabar\, r_g$, where we have defined $\lambdabar:=\lambda^2/(1+\lambda^2)\in(0,1)$,
and it describes the positive minimum of the area-radius function, i.e., $r_0\leq r$. The minimal surface $\mathcal{T}:=\{r=r_0\}$ (which \textit{always} lies inside the trapped region of spacetime because $0<\lambdabar<1$ and thus $r_0<r_g$)
is foliated by spheres of constant area $4\pi r_0^2$, and it splits the geodesically complete spacetime into two time-reversed regions (see Fig.~\ref{fig.diagram}). 
All curvature scalars are finite everywhere, and the spacetime is free of singularities.
For instance, the Ricci scalar and the Weyl Newman-Penrose scalars read, respectively,
\begin{align}\label{eq:scalarR}
R&=\frac{3r_g r_0}{2r^4},\\\label{eq:scalarpsi}
\Psi_2&=-\frac{3r_g}{2r^3}+\frac{r_0}{8r^4}(5r_g-2r).
\end{align}
It is important to note that,
since the model is explicitly covariant by construction, the length scales $r_g$ and $r_0$ are independent of any coordinate choice.

In the limit $r_0\to 0$, the theory reduces to GR and thus, the line element~\eqref{eq.metric} transforms into
the Schwarzschild geometry. In this limit, the surface $\mathcal{T}$ is replaced by a spacelike singularity, and
the maximal extension breaks into disconnected Kruskal regions.
Since the radius of the horizon $r_g$ remains unchanged in this limit, whenever we compare any of the regular geometries $r_0>0$ with Schwarzschild ($r_0\to 0$), we tacitly assume black holes with the same $r_g$. Nevertheless, other comparisons would also be possible. For instance, one could compare the regular geometry with a Schwarzschild black hole assuming they have the same asymptotic mass or surface gravity. We refer the interested reader to App.~\ref{app.others} for some examples.

For later use, we introduce the tortoise coordinate
for the exterior region of the horizon $r>r_g$ (gray shaded in Fig.~\ref{fig.diagram}),
\begin{align}\label{eq.tortoise}
    r_\ast&:=\int\sqrt{\frac{g_{rr}}{-g_{tt}}}\mbox{d}r=
    \int\frac{\mbox{d}r}{\sqrt{1-\frac{r_0}{r}}\left(1-\frac{r_g}{r}\right)},
\end{align}
where $g_{tt}$ and $g_{rr}$ stand for components of the metric in the diagonal chart \eqref{eq.metric}.
In fact, this integral can be explicitly performed to obtain an analytic expression for the tortoise
coordinate \cite{Alonso-Bardaji:2021yls,Alonso-Bardaji:2022ear},
    \begin{align}\label{eq.tortoiseexplicit}
r_*&=\frac{2r_g^{3/2}}{\sqrt{r_g-r_0}}
                    \Bigg[\ln\left(\frac{r}{{r_g}}-1\right)
                    -2\ln\left(\sqrt{\frac{r}{r_0}-1}
                    +\sqrt{1-\frac{r_0}{r_g}}\sqrt{\frac{r}{r_0}}\right)
                    \Bigg]\nonumber\\
              &+2\sqrt{r}\sqrt{r-r_0}
                +2r_g\left(2 + \frac{r_0}{r_g}\right)\ln\left(\sqrt{\frac{r}{r_0}}+\sqrt{\frac{r}{r_0}-1}\right),
\end{align}
which reduces to its standard form in GR in the limit $r_0\to 0$,
\begin{equation}
r_*|_{r_0\to0}=r+r_g \ln\left(\frac{r}{r_g}-1\right).
\end{equation}
In terms of this coordinate, the line element \eqref{eq.metric} takes the form
\begin{align}\label{eq.metrictortoise}
    \d s^2=g_{tt}\big(\d t^2-\d r_*^2\big)+r^2(r_*)(\d\vartheta^2+\sin^2(\vartheta) \d\varphi^2),
\end{align}
with $g_{tt}=-(1-\frac{r_g}{r})$.
Although the dependence on $r_0$ is not longer explicit in the expression of the metric,
the tortoise coordinate \eqref{eq.tortoiseexplicit} itself strongly depends on $r_0$.
However, qualitatively, the tortoise coordinate
has similar properties for all $r_0\geq 0$. In particular,
$r_*(r)$ is an asymptotically increasing function of $r$, such that
$r_*\to-\infty$ at the horizon $r\to r_g$, while it diverges as $r_*\to\infty$
as $r\to\infty$.
In addition,
from \eqref{eq.metrictortoise}, it is straightforward to see that
null radial geodesics satisfy $\d r_*/\d t=\pm 1$, and thus they are given
by $t\pm r_*={\rm constant}$. In this way, it is natural to define
the null outgoing ($v:=t+r_*$) and ingoing ($u:=t-r_*$) coordinates.

Below we will analyze the dynamics of a massless scalar test field propagating on
this background. Its corresponding potential will involve several terms but, in particular,
the derivative $\d^2r/\d r_*^2$. For any static, spherically symmetric
line element written as \eqref{eq.metrictortoise}, this derivative
can be generically rewritten in terms of the curvature of the manifold as follows,
\begin{align}\label{eq.tortoiserelation}
\frac{1}{r}\frac{\d^2r}{\d r_*^2}=g_{tt}\left(2\Psi_2+\frac{R}{6}\right),   
\end{align}
with $\Psi_2$ being the Coulomb term of the Weyl scalars \cite{szekeres1965gravitational}, 
$R$ the Ricci scalar, and $g_{tt}$ the corresponding metric component in \eqref{eq.metrictortoise}. 
For the present model, these can be identified in \eqref{eq:scalarR}, \eqref{eq:scalarpsi}, and \eqref{eq.metrictortoise}, respectively.

\subsection{Radiation of the horizon and dynamics of a massless scalar test field}\label{sec.radiation}

In the language of thermodynamics, any radiating body is characterized by its absorptivity, reflectivity, and transmittivity. A black-body radiator features a perfect absorption. If, for instance, the absorption were not perfect, one would speak of a gray body, which admits a gray-body factor that is defined through a nonvanishing transmittivity \cite{page1976particle,rosato2024ringdown}.

As it is well known, and it was first shown by Hawking \cite{hawking1975particle}, black-hole horizons emit a thermal spectrum due to particle production. An observer at infinity, who measures the radiation coming from the black hole, will perceive the spectral radiance (power per unit solid angle and unit projected area)
\begin{equation}\label{eq:bhemission}
    \int_0^\infty \d\omega\,\frac{n_p\hbar\omega^3}{8\pi^{3}}\Gamma_{\rm H}=\int_0^\infty \d\omega\,\frac{n_p\hbar\omega^3}{8\pi^{3}}\frac{\sigma(\omega)}{e^{\frac{E}{T}}\pm1},
\end{equation}
where $\omega$ is the frequency, $E=\hbar\omega$ the energy of the quanta, $T$ the temperature of the horizon, $\sigma(\omega)$ the gray-body factor, $n_p$ the number of linearly independent polarizations, and the $\pm$ accounts for bosons ($-$) and fermions ($+$), respectively. The radiance is composed out of the emission rate $\Gamma_{\rm BH}$, as well as the density of particle-states (per unit frequency and unit volume $\omega^2$) multiplied by the energy $\hbar\omega$ of one quantum, which combines to
$n_p \frac{\hbar\omega^3}{8\pi^{3}}$. For the sake of simplicity, we will conduct our analysis for bosons.  Since the density of states is unrelated to the specific geometric details, we will additionally restrict our analysis to the emission rate $\Gamma_{\rm H}$. Notice that the emission rate $\Gamma_{\rm H}$ further factorizes 
\begin{equation}
 \Gamma_{\rm H}=\frac{\sigma(\omega)}{e^{\frac{E}{T}}-1},
\end{equation}
into the Bose-Einstein distribution $1/(e^{E/T}-1)$, which describes the bosonic emission rate from a hot black body,
and a non-thermal part $\sigma(\omega)$ called the gray-body factor.
The Bose-Einstein distribution characterizes the Hawking effect at the horizon,
while the gray-body factor $\sigma(\omega)\in[0,1]$ describes the scattering of the modes from the gravitational potential,
that is, the transmittivity.

For the subsequent analysis, we will consider a massless,
scalar test field $\phi$,
which obeys the Klein-Gordon equation minimally coupled to the metric \eqref{eq.metric}, 
\begin{equation}\label{eq:KG}
    \Box\phi=-\frac{r}{r-r_g}\frac{\partial^2\phi}{\partial t^2}+\frac{1}{r^2}\frac{\partial}{\partial r}\left((r-r_g)(r-r_0)\frac{\partial\phi}{\partial r}\right)+\frac{1}{r^2}\Delta_\sphericalangle\phi=0,
\end{equation}
where $\Delta_\sphericalangle$ denotes the angular Laplace-Beltrami operator. 
To solve this equation, we will consider two different methods. Since the black-body part of $\Gamma_{\rm H}$ is solely determined by the horizon temperature, we employ the the tunneling method in Sec.~\ref{sec:tunneling}. This consists in a WKB (Wenzel-Kramers-Brillouin)
approximated solution to \eqref{eq:KG} evaluated in a near-horizon approximation.
For the gray-body factor, in Sec. \ref{sec:gbf}, we perform a mode decomposition
and consider appropriate boundary conditions for the field.

\section{Temperature of the black hole (Hawking radiation)}\label{sec:tunneling}

As mentioned above, the black-body part of the spectrum is determined through the temperature alone and marks the starting point for our analysis. 
To calculate the temperature for this black hole, we will work in the Hamilton-Jacobi formalism of the tunneling picture \cite{Parikh:1999mf,Shankaranarayanan:2000gb,Massar:1999wg}, which has been thoroughly studied
in Refs.~\cite{Vanzo:2011wq,Vanzo:2011nd,DiCriscienzo:2007pcr,DiCriscienzo:2010vz,Hayward:2008jq}. This method provides a quasi-local description of the Hawking effect and has proven itself to be extraordinarily versatile in various scenarios, e.g., for general dynamical horizons \cite{Giavoni:2020gui,senovilla2015particle} or modified dispersion relations \cite{DelPorro:2024tuw}. In fact, it can be shown that this method is
equivalent to the Bogolyubov approach \cite{unruh1976notes,israel1976thermo}, because the temperature originates from the non-analyticity at the horizon (cf. \cite{moretti2012state,DelPorro:2024tuw} for details on the equivalence).

From a technical perspective, the tunneling method uses the WKB formalism in which the field is represented by
\begin{equation}\label{eq:wkb}
    \phi=A\, e^{\frac i\hbar S_0},
\end{equation}
where $A$ denotes a slowly varying amplitude, that we treat as effectively constant,
and $S_0$ is the classical action (see Ref.~\cite{Vanzo:2011wq} for a detailed review of the method).
The ansatz \eqref{eq:wkb} should solve the equation of motion for the massless scalar field \eqref{eq:KG}. To ensure that this is the case,
we define the momenta $k_a=\nabla_aS_0$, such that
\begin{equation}\label{eq:snull}
    S_0=\int_\mathcal{M} k_a\mbox{d}x^a=-\int E\mbox{d}t+\int k\mbox{d}r+W(\sphericalangle),
\end{equation}
with $E:=-\mathcal{L}_\xi S_0$ being the energy with respect to the timelike Killing vector $\xi$, $k$ the radial momentum, and $W(\sphericalangle)$ the angular contribution. We will work with s-waves, that is, $W(\sphericalangle)\equiv0$ in the remainder. Due to the symmetries of the system, this is the most probable form of radiation \cite{Vanzo:2011wq}.

As in quantum mechanics, the tunneling rate is determined by a comparison between the incident and the transmitted intensity \cite{Vanzo:2011wq}
\begin{equation}\label{eq:tunnelrate}
    \Gamma=\frac{|\phi_{\rm trans}|^2}{|\phi_{\rm inc}|^2}=e^{-\frac{2}{\hbar}{\rm Im}(S_0)}.
\end{equation}
This rate is \emph{a priori} not necessarily related to any thermal process. To understand the Hawking effect in this framework, we compare the tunneling rate with the thermal Boltzmann distribution. If $\Gamma\propto e^{-E/T}$, then we would perceive a thermal radiation, namely the Hawking effect. That is, provided that Im$(S_0)\propto E$, the Hawking effect describes the process for which \cite{Giavoni:2020gui}
\begin{equation}
    \mbox{Im}(S_0)>0.\label{eq:defHE}
\end{equation}
This is intuitively clear since, in this case,
the rate assumes a thermal distribution and, thus, a positive-definite imaginary part that
leads to a positive-definite horizon temperature $T$. 

By using \eqref{eq:wkb} with \eqref{eq:snull} in \eqref{eq:KG}, we transform the Klein-Gordon differential equation into an algebraic equation for the $k_a$.
Since we have chosen $E$ to be associated with the Killing vector
field, it is a conserved quantity, such that the $t$-integration in \eqref{eq:snull}
yields no imaginary part. Therefore, it suffices to focus on $k$. In our example, we find for $k$ the outgoing $k_-$ and ingoing $k_+$ solutions
\begin{equation}\label{eq:momenta}
k_\pm=\mp\frac{E}{\left(1-\frac{r_g}{r}\right)\sqrt{1-\frac{r_0}{r}}},%
\end{equation}
which \emph{a priori} does not necessarily lead to imaginary contributions. In fact, an imaginary part can only occur when $k$ admits a simple pole at the horizon.
In the present case, as will be explained below,
$k$ develops an imaginary contribution by standard arguments from distribution theory and complex analysis \cite{hormander2002analysis}.
For definiteness, since we are interested in the outward tunneling, we choose for
the remainder $k_-$ as the solution. 

The form \eqref{eq:momenta} already shows a simple pole, which allows us to perform the integration immediately after introducing a small complexification $r-r_g\to r-r_g+i0$ around this simple pole. In this way, we get,
\begin{equation}\label{eq:ims0}
    \mbox{Im}(S_0)=\mathrm{Im}\left(\int\frac{rE\mbox{d}r}{\sqrt{1-\frac{r_0}{r}}(r-r_g+i0)}\right)=\frac{2r_g\pi E}{\sqrt{1-\frac{r_0}{r_g}}}.
\end{equation}
Here we used a result from distribution theory, called Sokhotski-Plemelj theorem in its integral form, which is based on \cite{hormander2002analysis} 
\begin{equation}
\frac{1}{y\pm i0}=\mp i\pi\delta(y)+\frac{1}{y_+}-\frac{1}{y_-},
\end{equation}
where we defined the homogeneous distributions $y_\pm$ such that $y_+=y$ whenever $y>0$ and $y_-=|y|$ when $y<0$, and otherwise both are zero\footnote{When this extension of the homogeneous distribution is integrated against a function $f(y)\in\mathcal{C}^1_0(\mathbb{R})$, the last two terms combine to the Cauchy principal value
\begin{equation}
   \mathrm{PV}\left(\int \frac{f(x)}{x}\mbox{d}x\right)= \lim_{\varepsilon\to0}\int_{|x|>\varepsilon}\frac{f(x)}{x}\mbox{d}x.
\end{equation}}.
Comparing with the definition \eqref{eq:defHE}, we can see that our imaginary part is positive definite and proportional to $E$ and, thus, constitutes a thermal radiation, that is,
the Hawking effect. The corresponding temperature is derived by comparing the tunneling rate \eqref{eq:tunnelrate} with a Boltzmann distribution, such that, taking into account Eq. \eqref{eq:ims0}, we read off the horizon temperature 
\begin{equation}\label{eq:hawkingtemp}
    T=\frac{\hbar E}{2\mbox{Im}(S_0)}=\frac{\hbar}{4\pi r_g}\sqrt{1-\frac{r_0}{r_g}}.
\end{equation}
Generically \cite{cropp2013surface}, the temperature is directly related
to the surface gravity\footnote{The surface gravity is defined in terms of the norm squared of the gradient of the Killing
field as $\kappa^2:=-g(\nabla\xi,\nabla\xi)/2|_{\hor}=(r_g-r_0)/(4 r_g r_0^2)$.} $\kappa$
at the horizon as $T=\hbar\kappa/(2\pi)$.
As expected, the temperature depends on both parameters of the theory,
$r_0$ and $r_g$, and for $r_0\to0$ one recovers the standard expression corresponding to Schwarzschild black holes.
For a horizon of a given fixed size $r_g$, its temperature decreases with $r_0$, being maximal for the
Schwarzschild black hole $(r_0\to0)$ and minimal for the limit $r_0\to r_g$.
Therefore, $r_0$ causes a decreasing temperature as compared with the Schwarzschild case.
In fact, in the upper limit of $r_0$, i.e., $r_0\to r_g$, the temperature
vanishes. Therefore, as will be explained below in more detail, if the black hole reaches this particular limit,
the radiation will cease and, instead of completely evaporating, the black hole will leave
behind a stable (nonradiative) remnant of size $r_0$.

Once the temperature of the horizon, which completely characterizes the black-body spectrum
of the horizon, has been obtained, we continue to analyze the gray-body factor.

\section{gray-body factor (potential scattering)}\label{sec:gbf}

In this part, we refine our analysis regarding the phenomenology of the
radiation from the black hole by deriving the gray-body factor $\sigma(\omega)$ defined in Eq. \eqref{eq:bhemission}. This section is divided into two subsections. In Sec.~\ref{sub:msd}
we perform a mode decomposition of the scalar field and obtain the Regge-Wheeler equation.
This equation can be understood as a Schr\"odinger equation for a particle in a potential.
Then, in Sec.~\ref{sub:bc}, we introduce the corresponding boundary conditions
and construct the gray-body factor as the transmission coefficient of a scattering process
considering the mentioned potential.

\subsection{Mode decomposition and Regge-Wheeler equation}\label{sub:msd}

Using the isometries
of the metric, and recalling that $\partial_t$ is a Killing vector field, we perform a mode-sum decomposition for the field
\begin{equation}\label{eq:msd}
    \phi(t,r,\vartheta,\varphi)=\frac{1}{r}\int_\mathbb{R}\frac{\mbox{d}t}{2\pi}\sum_{l,m}\psi_l(r;\omega)e^{-i\omega t}Y_{lm}(\vartheta,\varphi),
\end{equation}
into spherical harmonics $Y_{lm}(\vartheta,\varphi)$ and Fourier modes $e^{-i\omega t}$. 
With equation \eqref{eq:msd}, we transform the partial differential equation \eqref{eq:KG}
for $\phi$ into an ordinary differential equation for the mode $\psi_l$, which we recast in a simpler form using the tortoise coordinate \eqref{eq.tortoise}. More precisely, since the spherical harmonics are the eigenfunctions of the angular Laplace-Beltrami operator $\Delta_\sphericalangle Y_{lm}(\vartheta,\varphi)=-l(l+1)Y_{lm}(\vartheta,\varphi)$, we find the Regge-Wheeler equation:
\begin{equation}\label{eq:rwe}
    \frac{\mbox{d}^2\psi_l(r_\ast;\omega)}{\mbox{d}r_\ast^2}+\left(\omega^2-V_l(r(r_\ast))\right)\psi_l(r_\ast;\omega)=0.
\end{equation}
It follows that the modified Regge-Wheeler potential \cite{Alonso-Bardaji:2021yls} can be written in the general form
\begin{align}\label{eq:aform}
    V_l(r)&=-g_{tt}\frac{l(l+1)}{r^2}+\frac{1}{r}\frac{\d^2r}{\d r_*^2} 
 =-g_{tt}\left(\frac{l(l+1)}{r^2}-2\Psi_2-\frac{R}{6}\right)\nonumber\\
 &=\left(1-\frac{r_g}{r}\right)\left(\frac{l(l+1)}{r^2}+\frac{2r_g+r_0}{2r^3}-\frac{3r_gr_0}{2r^4}\right),
\end{align}
where we used \eqref{eq.tortoiserelation} in the last step. It is clear from the above equation that the limit $r_0\to 0$ yields the usual Regge-Wheeler potential in Schwarzschild spacetime.

The general form in \eqref{eq:aform} provides already some hints towards the interpretation and physical consequences of $r_0$. Let us focus on the second parentheses in the last equation. First, the centrifugal term, $l(l+1)/r^2$, does not acquire any corrections due to $r_0$.
Since it is the dominant term for large radii, for $l\neq0$ the asymptotic behavior of the potential is equal to the Schwarzschild case for any value of $r_0$. Second, the Weyl scalar $\Psi_2$ shifts the coefficient of $1/r^3$ slightly. Third, the sum of both $\Psi_2$ and the Ricci scalar $R$
leads to a negative contribution,
which is not present in the Schwarzschild case and it decays as $1/r^4$. As we will see, this last
term diminishes the height of the maximum of the potential, and thus we expect
to see a smaller value of the gray-body factor than for Schwarzschild.
\begin{figure}
    \centering
    \includegraphics[width=.9\textwidth]{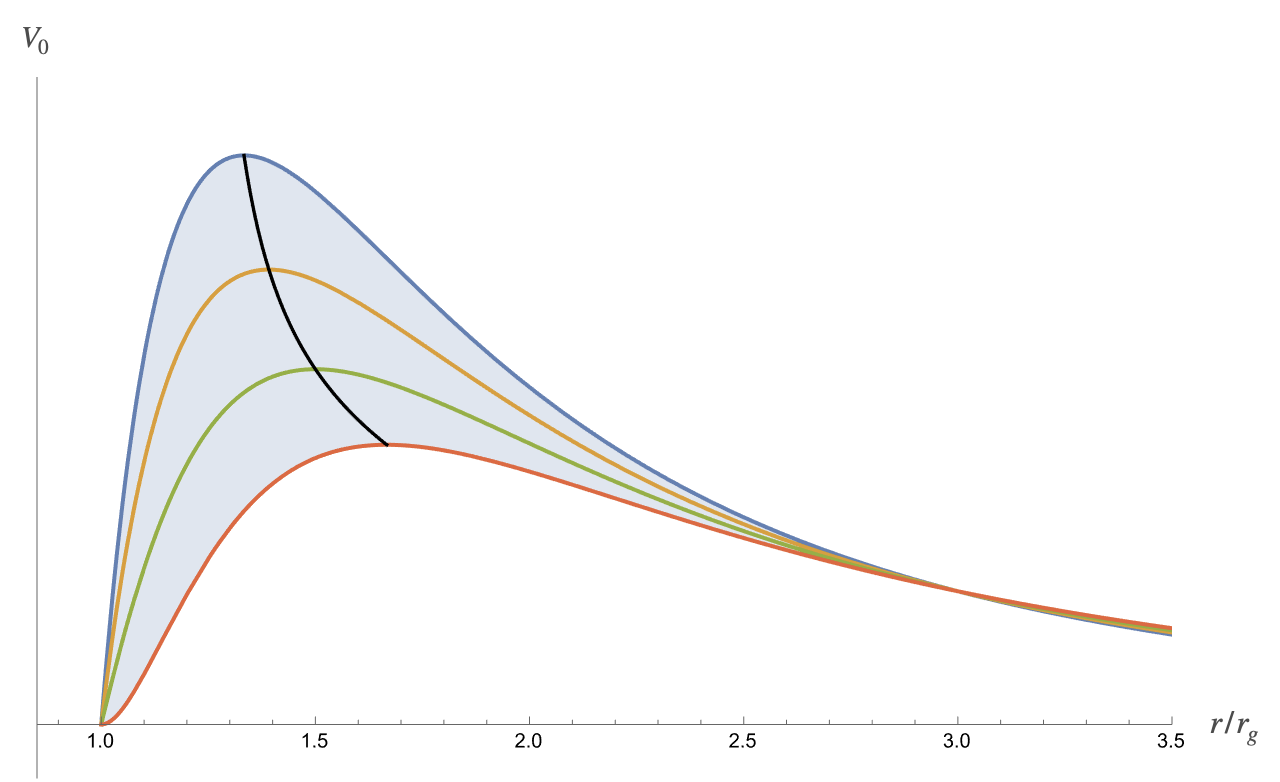}
    \caption{Regge-Wheeler potential with $l=0$ for black holes of same horizon-radius
    $r_g$. Each curve corresponds to a value $r_0/r_g \in\{0,1/3,2/3,1\}$ in $\{$blue, orange, green, red$\}$. The shadowed region represents the region covered by the corresponding potentials for all possible values of $r_0/r_g$. The black line joins together the maximum of the potential in each case. Note that, for any $r_0/r_g$,
    the potential takes the same values at $r=r_g$ and at $r=3r_g$.}
    \label{fig:rh}
\end{figure}

As can be seen in Fig.~\ref{fig:rh}, 
the shape of the potential $V_l(r)$ outside the horizon has the same qualitative form for any $r_0\in[0,r_g]$. 
It vanishes at $r=r_g$, then it increases until reaching a maximum at $r=r_{\rm max}$, which always lies between $r=4r_g/3$ and $r=5r_g/3$. More precisely, these boundary values correspond
to $\{l=0,r_0\to 0\}$ and to $\{l=0, r_0\to r_g\}$, respectively. For any value of $r_0$,
as we increase the value of the angular mode number $l$, the height of the potential
increases and the position of the maximum $r_{\rm max}$ tends to the value $3r_g/2$.
For more details on $r_{\rm max}$, we refer the reader to App.~\ref{app:rmax}.
 
When taking the difference between the potential with $r_0>0$ and the one corresponding to Schwarzschild $(r_0\to 0)$, it is possible to see that their difference is independent of the angular momentum,
\begin{align}\label{eq:rwpasy}
    V_l(r)-V_l(r)|_{r_0\to 0}=\frac{r_0}{2r^5}\left(r-{r_g}\right)\left(r-{3r_g}\right).
\end{align}
In addition, it is easy to see that this difference vanishes at $r=r_g$ and $r=3 r_g$,
such that the difference \eqref{eq:rwpasy} is negative for $r_g<r<3r_g$ and positive for $3r_g<r$.
Therefore, as commented above, the parameter $r_0>0$ lowers the height of the
maximum of the potential as compared to the Schwarzschild case
(because it is always located in the interval $4r_g/3\leq r_{\rm max}\leq 5r_g/3$),
and it induces a slight displacement of $r_{\rm max}$ toward larger values of $r$. 
Asymptotically, the effect of $r_0$ turns into the contrary, therefore, it increases the value of the potential. However, this last effect can only be appreciated in s-waves ($l=0$),
because for $l\neq0$ the centrifugal term dominates at large distances
and the contributions from $r_0$ are negligible. 
Recall that, for the comparison with Schwarzschild, we are assuming black holes of equal `size' $r_g$. In App.~\ref{app.others} we compare black holes of equal ADM mass and equal temperature, for which the above conclusions are not necessarily true.

In any case,
concerning the asymptotics, $V_l(r)\to 0$ whenever the tortoise coordinate diverges $r_\ast\to\pm\infty$,
that is, both at the horizon and at infinity. This is true for any $r_0$, and thus the Regge-Wheeler potential always behaves exactly as in GR on its limits. In particular, from the vanishing of $V_l(r)$ at the horizon, it becomes a posteriori clear why the tunneling prescription followed in Sec. \ref{sec:tunneling} captured the relevant pieces, although it was restricted to s-waves.

\subsection{Boundary conditions and the gray-body factor as the transmission coefficient}\label{sub:bc}

\begin{figure}
    \centering
    \includegraphics[width=.7\textwidth]{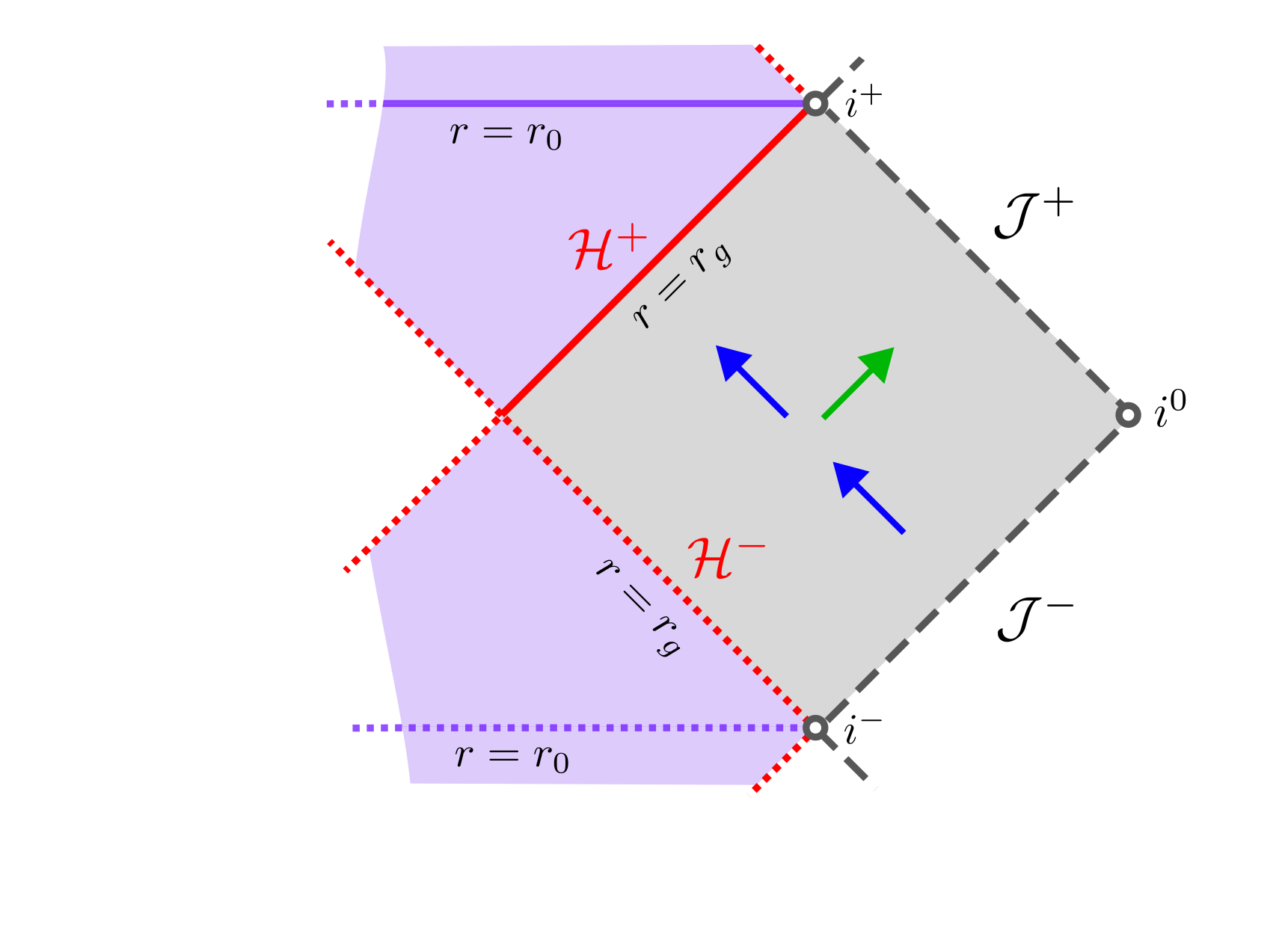}
    \caption{We illustrate the splitting of the modes in the external region (shaded in gray) of the black hole. The ingoing modes coming from past null infinity split into reflected upgoing (green) modes and transmitted ingoing (blue) modes that contribute to the gray-body factor.}
    \label{fig:modecom}
\end{figure}

The commented properties of the potential allow us to construct the modes in the past and future null infinities, $\mathcal{J}^-$ and $\mathcal{J}^+$, by ingoing and outgoing plane waves. In Fig.~\ref{fig:modecom} we illustrate the setup. We assume only ingoing modes $e^{-i\omega v}$ at $\mathcal{J}^-$. These modes face a dichotomous fate: they either fall through the horizon into the black hole, being described by $e^{-i\omega v}$ at $\mathcal{H}^+$, or they get reflected by the potential and turn into outgoing modes $e^{-i\omega u}$ that propagate to $\mathcal{J}^+$. Since the dependence on $t$ is only given by the exponential $e^{-i\omega t}$, we can factor this out as in the decomposition \eqref{eq:msd},
such that we are left with the usual boundary conditions at the horizon and at spatial infinity, respectively,
\begin{eqnarray}
    \psi_l(r_\ast;\omega)&\sim&T_l(\omega)e^{-i\omega r_\ast}, \qquad\hspace{0.97cm}\mbox{for}\quad r_\ast\to-\infty,\\
    \psi_l(r_\ast;\omega)&\sim&e^{-i\omega r_\ast}+R_l(\omega)e^{i\omega r_\ast}, \quad\mbox{for}\quad r_\ast\to+\infty.
\end{eqnarray}
Here, we defined the transmission and the reflection coefficients, $T_l(\omega)$ and $R_l(\omega)$, that fulfill the standard
normalization condition $|T_l(\omega)|^2+|R_l(\omega)|^2=1$ and are essential to determine the gray-body factor. Therefore, for $r_\ast\to+\infty$, we can find the ingoing modes, that were prepared in the past, together with the reflected outgoing modes, that arrive in the future. At the horizon, $r_*\to-\infty$, only infalling modes are present. Since the gray-body factor $\sigma_l\in[0,1]$ captures the scattered modes, its definition is given by 
\begin{equation}
    \sigma_l(\omega)=|T_l(\omega)|^2=1-|R_l(\omega)|^2.
\end{equation}
We note that, in the present and the following section, we will usually present
the explicit expression for the modulus squared of the reflection coefficient
$|R_l(\omega)|^2$, which, taking into account this last relation,
automatically provides the form of the gray-body factor.

In order to calculate $R_l(\omega)$ and $T_l(\omega)$ we follow the approach of Refs.~\cite{Fabbri:1975sa,Konoplya:2019hlu},
which employs the WKB approximation. Under consideration of the stationary phase
approximation---we work in a close neighborhood of the maximum of the potential---, we formally integrate \eqref{eq:rwe} to find
\begin{equation}\label{eq:wkbwave}
    \psi_l(r_\ast;\omega)\propto e^{\pm \int \sqrt{\bar\Omega_l(r(r_\ast);\omega)}\mathrm{d}r_\ast},
\end{equation}
and define the WKB frequency $\bar\Omega_l(r(r_\ast);\omega):=V_l(r(r_\ast))-\omega^2$. Note that
this formal solution complies with the asymptotic values because $\lim_{r_\ast\to\pm\infty}V_l(r(r_\ast))=0$ and, therefore, $\lim_{r_\ast\to\pm\infty}\bar\Omega_l(r(r_\ast);\omega)=-\omega^2$.

\begin{figure}
    \centering
    \includegraphics[width=.8\textwidth]{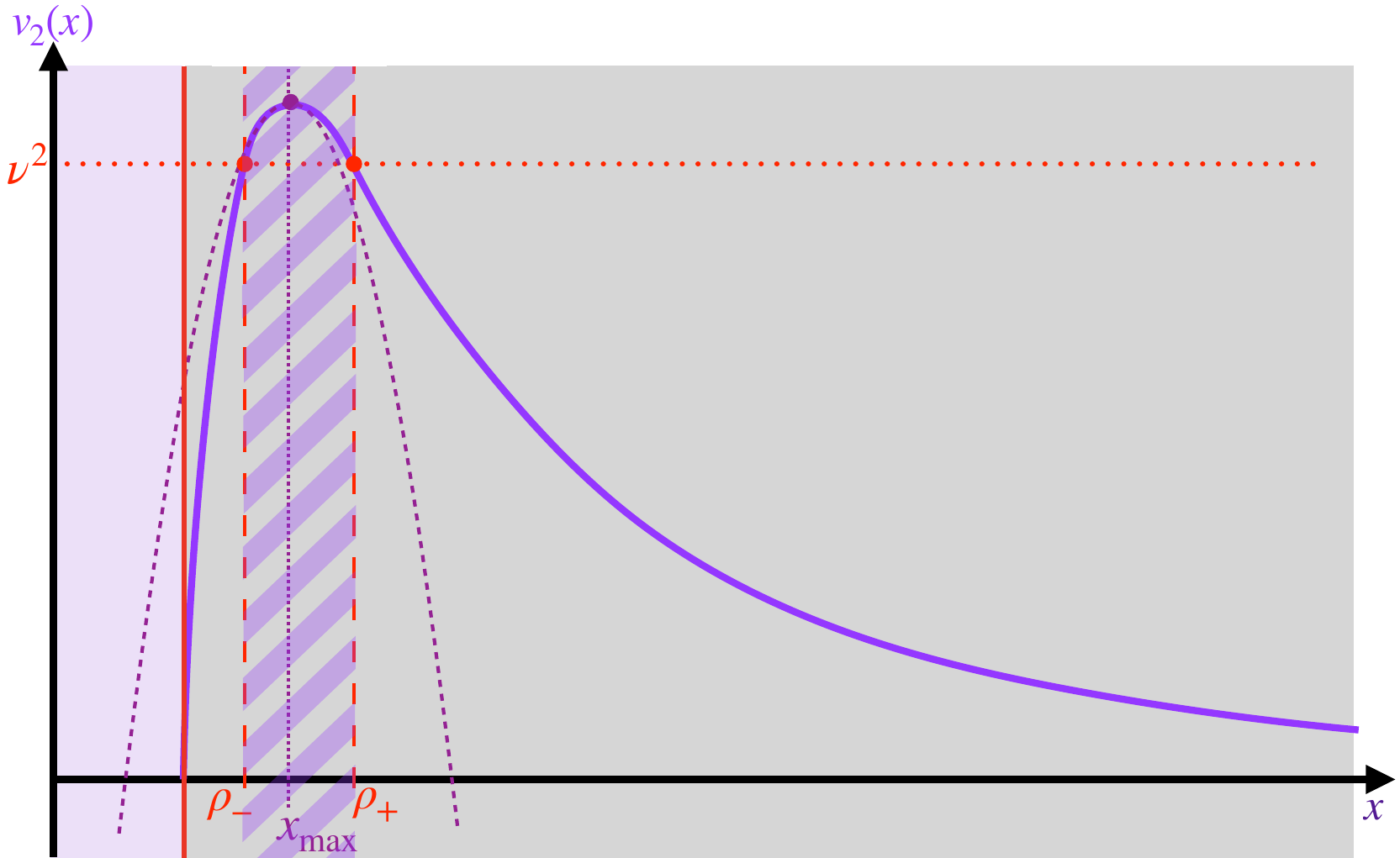}
    \caption{Regge-Wheeler potential for $l=2$ to illustrate the different regions and points occurring in the WKB analysis. The color scheme matches those in the Penrose diagrams, where gray is the exterior region.
    In fact, the potential is only displayed for the exterior region. The red dotted line represents the chosen frequency $\nu^2$ and cuts the potential into three regions, limited by the turning points $\rho_\pm$. The intermediate region where $\nu^2\leq v_2(x)$ is highlighted through hachures. In the formalism, the potential is approximated (at the leading order around the maximum $x_{\rm max}$) by an inverted parabola represented by the dashed line.}
    \label{fig:rwleg}
\end{figure}

To apply the ansatz \eqref{eq:wkbwave}
for $\psi_l$, we need to choose an interval for $r_\ast$ such that either $\bar\Omega_l(r(r_\ast);\omega)>0$ or $\bar\Omega_l(r(r_\ast);\omega)<0$.
Then, to describe the full scattering process,
the next step is to match the $\psi_l$ at the turning points.
These turning points are given by the roots $r=\rho\, r_g$ that
solve $\bar\Omega_l(r(r_\ast);\omega)|_{r=\rho \,r_g}=0$. 
Due to the $r_0$-dependent terms, this equation implies finding the roots of a fifth-order
polynomial, instead of a fourth-order polynomial like in the Schwarzschild spacetime.
However, as can be clearly seen in Fig.~\ref{fig:rwleg},
there are at most two turning points outside the horizon, which imply
two real roots, $\rho_{+}$ and $\rho_{-}$, such that $1<\rho_{-}\leq\rho_{+}$.

For the particular form of the commented fifth-order polynomial,
there exist techniques to determine $\rho_+$ and $\rho_-$ analytically (cf. \cite{dummit1991solving,lazard2004solving}),
but these yield complicated expressions that allow to extract only limited information about the system. Instead, to develop the phenomenology analytically, we follow Konoplya \cite{Konoplya:2019hlu}, who uses the same method as in Refs.~\cite{schutz1985black, Iyer:1986np}, based on an expansion around the maximum of the potential.

For the sake of technical simplicity, we will work with $r(r_\ast)$ from now on, if not stated otherwise. Additionally, let us introduce the set of dimensionless quantities $x:=\frac{r}{r_g}$ and $\nu:=r_g\,\omega$, as well as $\xzero =\frac{r_0}{r_g}$, which was already defined
above. These definitions change the WKB frequency $\bar\Omega_l(r;\omega)$ to
\begin{equation}\label{eq:Omegabar}
    \Omega_l(x;\nu):=r_g^2\,\bar\Omega_l(x;\nu)=v_l(x)-\nu^2,
\end{equation}
where
the dimensionless potential is given by
\begin{align}\label{eq:dimlospot}
    v_l(x):=r_g^2 V_l(x)=\left(1-\frac{1}{x}\right)\left(\frac{l(l+1)}{x^2}+\frac{2+\xzero}{2x^3}-\frac{3\xzero}{2x^4}\right).
\end{align}
Since the WKB approximation connects the asymptotic conditions with the behavior at the maximum of the potential barrier through the turning points, we can relate ingoing and outgoing modes through an S-matrix like description.

On the one hand, for large enough frequencies, $v_l(x_{\rm max})<\nu^2$, there exist no turning points, and
the potential is effectively transparent for all the modes. Thus the gray-body factor $\sigma_l(\nu)$
approximates one and we are left with a black-body spectrum.

On the other hand, for relatively small frequencies, such that $\nu^2<v_l(x_{\rm max})$,
there are two roots outside the horizon, and
we divide our system into three regions: $x<\rho_-$, $\rho_-\leq x\leq \rho_+$, and $\rho_+<x$.
For the first and the third regions, the frequency squared is larger than
the potential, $\nu^2>v_l(x)$, and thus the particle moves almost freely.
In the intermediate region, $\rho_-\leq x\leq \rho_+$, the frequency squared is smaller than (or equal to) the potential,
$\nu^2\leq v_l(x)$, and the dynamics is heavily influenced by the potential.
In this region, the WKB frequency $\Omega_l$ is approximated by an expansion around the maximum $x_{\rm max}$,
\begin{equation}
    \Omega_l(x;\nu)=\Omega_l(x_{\rm max};\nu)+\frac{\d^2\Omega_l}{\d x^2}(x_{\rm max};\nu)(x-x_{\rm max})^2+\mathcal{O}((x-x_{\rm max})^3).
\end{equation}
In general, the matching at the turning points turns out to be cumbersome, but a good result can be achieved within the eikonal approximation, which effectively truncates the expansion at the second order, such that the potential barrier is approximated by a downward facing parabola (see Fig.~\ref{fig:rwleg}).
For this approximation to be valid,
for any $x$ in the intermediate region, $\rho_-\leq x\leq \rho_+$, the distance to the maximum should not exceed
\begin{equation}\label{eq:cond}
    \max_{\rho_-\leq x\leq \rho_+}|x-x_{\rm max}|=(\rho_+-x_{\rm max})<\sqrt{\frac{-2\Omega_l(x_{\rm max},\nu)}{\frac{\d^2\Omega_l}{\d x^2}(x_{\rm max},\nu)}}.
\end{equation}
The equality in the last equation comes from the fact that $(x_{\rm max}-\rho_-)<(\rho_+-x_{\rm max})$,
which is straightforward to see from the slope of the curve. 
This marks the validity of the Taylor approximation to quadratic order in the potential. By a standard analysis
in black-hole perturbation theory \cite{Konoplya:2019hlu}, one finds, within the eikonal approximation,
that the WKB frequency in the intermediate region is given by
$\Omega_l(x;\nu)\approx K_l^2 (\nu)$, where
\begin{equation}\label{eq:kl}
    K_l(\nu):=-i\frac{ \Omega_l(x_{\rm max},\nu)}{\sqrt{-2\frac{\d^2\Omega_l}{\d x^2}(x_{\rm max},\nu)}}.
\end{equation}
Having determined the WKB solution, one can obtain the reflection and transmission coefficients
\cite{Iyer:1986np,Konoplya:2019hlu},
\begin{eqnarray}
    |R_l(\nu)|^2&=&\frac{1}{1+e^{-2\pi i K_l(\nu)}},\label{eq:reflexion}\\
    |T_l(\nu)|^2&=&\frac{1}{1+e^{2\pi i K_l(\nu)}}.\label{eq:transmission}
\end{eqnarray}
We see that the coefficients fulfill the normalization condition, and solely depend on the evaluation of the WKB frequency
and its second-order derivative at the maximum $x_{\rm max}$. Refinements of this treatment can be achieved
via Pad\'e approximants or considering higher-orders terms in the Taylor expansion \cite{Konoplya:2019hlu}.
With this analysis we conclude the general computation of the
gray-body factor. However, in order to extract specific properties, in the next section we will discuss
the main features of the black-hole radiation during different evolutionary epochs: from an astrophysical
black hole towards a remnant.

\section{Radiation from the nonsingular black hole during different evolutionary epochs}\label{sec:examples}

To chart the thermal history of the nonsingular black hole, we study the two extreme states: the large, astrophysical black hole and the almost-remnant configuration. This section
is divided into two subsections. In Sec.~\ref{sec:apbh} we discuss the gray-body factor of astrophysical black holes, first for large angular mode number $l$ and then the particular case of s-waves ($l=0$), where the centrifugal barrier vanishes. In this latter case, the effects of the parameter $r_0$ become a prominent feature of the potential.
In Sec.~\ref{sec:arc} we assume the evaporation to be almost completed,
and study the gray-body factor at the moment right before the remnant state is established.
In this case, we also consider the large-$l$ and the s-waves ($l=0$) limits.

\subsection{Astrophysical black hole}\label{sec:apbh}

Astrophysical black holes describe the stage that is dominated by classical gravity
and should feature small quantum-gravity corrections.
This corresponds to relatively large black holes that have not yet radiated away much of their initial mass. As such, the transition surface, located at $r=r_0$, resides far away from the horizon at $r=r_g$. This corresponds to the limit $\xzero\ll1$.

\subsubsection{Large $l$ analysis}\label{sec.apbhll}

We follow the analysis in Ref.~\cite{Fabbri:1975sa} and first determine the gray-body factor for large angular mode number $l$.
In this limit, our second-order approximation should be especially good because the higher $l$ becomes, the more the height of the potential increases, which in turn allows for larger frequencies. As a first step, we express the position of the maximum, given in \eqref{eq:max}, in dimensionless quantities and expand for large $l$ to
obtain an approximate expression for $x_{\rm max}$,
\begin{equation}\label{eq:max1}
    x_{\rm max}=\frac32+\frac{3\xzero -2}{12\lk}+\mathcal{O}\left(\frac{1}{(\lk)^{2}}\right).
\end{equation}
A comparison with the approximate turning points in \cite{Fabbri:1975sa} assures that the maximum can be found around the photosphere, which is located at $x=\frac32$ in our unit system. This position is well located between the approximate turning points for Schwarzschild,
\begin{equation}\label{eq:wepu}
    \rho_\pm\Big|_{\lambdabar\to 0}=\frac{3}{2}\left(1\pm\frac{1}{\sqrt{3}}\sqrt{1-\frac{27\nu^2}{4l(l+1)}}\right)+\mathcal{O}\left(\frac{1}{\sqrt{(\lk)^3}}\right).
\end{equation}
When using our maximum \eqref{eq:max1} to construct $K_l(\nu)$, as defined in Eq. \eqref{eq:kl},
we expand the resulting expression in terms of large $l$ to find the simple formula
\begin{align}\label{eq:expKLbig}
 K_l(\nu)= \;-i\frac{\sqrt{\lk}}{6}+i\left(\frac{9 \nu^2}{8}-\frac{\xzero }{27}\right)\frac{1}{\sqrt{\lk}}+\mathcal{O}\left(\frac{1}{\sqrt{(\lk)^3}}\right),
\end{align}
which we can feed into the expression for the reflection coefficient \eqref{eq:reflexion}.
We note that the $\xzero $-dependent contributions carry a different sign than the other terms of the same order in $l$, and, therefore, should reduce the reflected contribution. By using \eqref{eq:expKLbig} in \eqref{eq:reflexion} we find the small-$\xzero$ corrections
\begin{equation}\label{eq:rapprox}
    |R_l(\nu)|^2=\frac{1}{f_l^o(\nu)+1}+\frac{\pi  \xzero \left(8 (\lk+3)-135 \nu^2\right)f_l^c(\nu)}{108 (\lk)^{3/2} \left(f_l^c(\nu)+e^{\frac{9 \pi  \nu^2}{4 \sqrt{\lk}}}\right)^2}+\mathcal{O}(\xzero^2),
\end{equation}
where we defined the two quantities at the respective orders: one in the
$\lambdabar$-independent contribution $f_l^o(\nu)$ and one in the correction term $ f_l^c(\nu)$,
\begin{eqnarray}
    f_l^o(\nu)&:=&\exp \left(\frac{\pi  \left(243 \nu^2 (3 \lk-2)-108 (\lk)^2+32\right)}{324 (\lk)^{3/2}}\right),\label{eq:upso}\\
    f_l^c(\nu)&:=&\exp \left(\frac{\pi  \left(729 \nu^2 (6 \lk-1)+4 \left(243 (\lk)^3-72 \lk+32\right)\right)}{2916 (\lk)^{5/2}}\right)\label{eq:upsc}.
\end{eqnarray}
From \eqref{eq:rapprox}, we observe that the $\nu$-dependence in the correction term comes with a negative sign, implying that, for $\nu>\frac{8\lk+24}{135}$, the correction acts subtractive. In general, for this treatment to remain sensible, $\nu$ cannot take arbitrary values. More precisely, for large $l$, condition \eqref{eq:cond} requires 
\begin{equation}
   \nu>\frac29\sqrt{ \lk \left(6 - 4 \sqrt{3} |\rho_+-x_{\rm max}|\right)-2  +\frac{7 \xzero}{3} },
\end{equation}
therefore suggesting that higher $l$ require higher frequencies for this analysis to remain valid. In other words, in this limit the turning points remain close enough to the maximum,
which can only be achieved for large enough frequencies. 

If one leaves these limits, the eikonal-approximated formula may not be sufficiently accurate and we would need to add additional contributions to gain meaningful results (cf. \cite{Konoplya:2019hlu} for details). 
\begin{figure}
    \centering
    \includegraphics[width=0.47\textwidth]{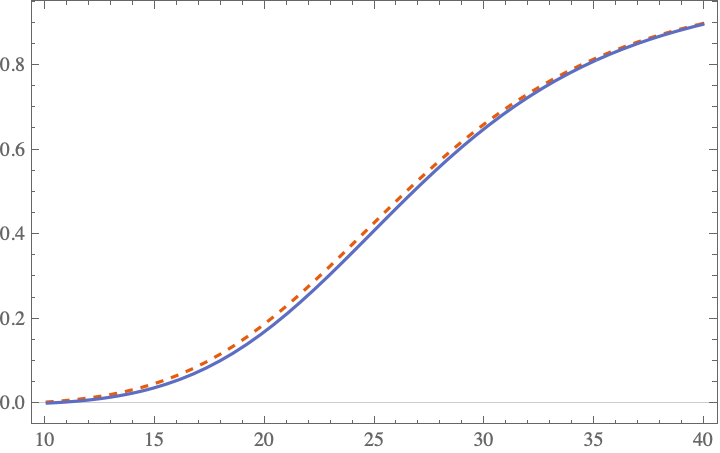}
     \includegraphics[width=0.475\textwidth]{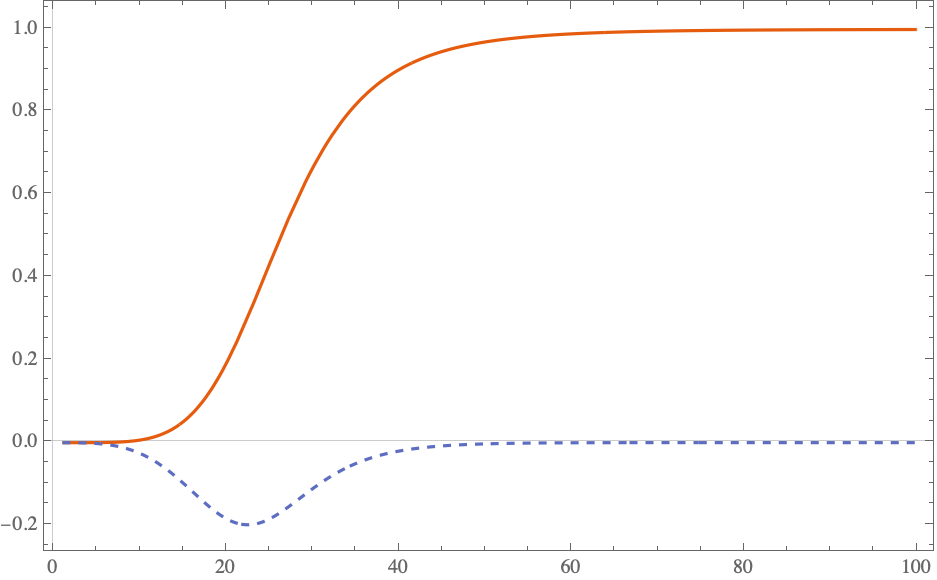}
    \caption{For $\nu=2$, we plotted the reflection coefficient $|R_l|^2$ as a function of $l$. Left panel: we see the reflection coefficient for $\xzero\in\{0,0.9\}$ to illustrate the influence of $\lambdabar$. The dashed line corresponds to $\xzero=0$, while the bold line shows the value $\lambdabar=0.9$. We clearly see a slight downwards shift of the reflection. The right panel displays the two individual contributions in $|R_l|^2$, that is, the $\xzero$-dependent (bold line) and $\xzero$-independent (dashed line) term. It is obvious that the correction lowers the original ($\xzero=0$) contribution only $l$-locally around the inflection point. Note that we multiplied the correction of $|R_l|^2$ (dashed blue line) by a factor $1000$ to make the correction visible in the plot; for the actual value of $\xzero=0.01$, the minimum would be around the value $-2\cdot10^{-4}$.}
    \label{fig:reflectionforx0}
\end{figure}
In Fig.~\ref{fig:reflectionforx0}, we plotted on the left panel the reflection coefficient
for the frequency $\nu=2$ against $l$ for two particular values of $\xzero$.
As expected, the $\xzero $-contribution lowers the reflection coefficient, which implies a larger gray-body factor.
In fact, its modification is most palpable around the inflection point of $|R_l(\nu)|^2$
as a function of $l$. 

From the complete black-hole evaporation rate \eqref{eq:bhemission}, we derive the rate for every $l$-mode to be
\begin{equation}\label{eq:emmisionsrate}
    \Gamma_{\rm H}^{l}=\int\frac{\sigma_l(\nu)}{e^{\frac{\nu}{\theta}}-1}\d\nu=
    \Gamma_{\rm H}^{l}\Big|_{\lambdabar\to 0}+\xzero\gamma_l,
\end{equation}
where the dimensionless horizon temperature is defined as $\theta:=r_gT/\hbar=\sqrt{1-\xzero }/(4\pi)$. We see that the integral splits into a $\xzero$-independent part and a part $\gamma_l$ from the corrections. Since $\xzero$ shows up in both the black-body as well as the gray-body contribution, $\gamma_l$ can be split as $\gamma_l=\gamma_l^{\rm black}+\gamma_l^{\rm gray}$. On the one hand, the black-body contribution $\gamma_l^{\rm black}$
can be derived with \eqref{eq:hawkingtemp} and \eqref{eq:rapprox} to be
\begin{equation}
   \gamma_l^{\rm black}= -\frac{\theta}{2}\int\frac{e^{\frac{\nu}{\theta}}\left(1-|R_l(\nu)|^2\right)}{\left(e^{\frac{\nu}{\theta}}-1\right)^2}\Bigg|_{\xzero=0}\d\nu.
\end{equation}
The minus sign in front of this term would reduce the value of the integral,
and thus of $\Gamma_H^l$. This feature can be understood from \eqref{eq:emmisionsrate},
because larger values of $\xzero$ decrease the temperature $\theta$. Due to the inverse dependence,
the exponent increases and induces a higher suppression coming from the Planck distribution.
On the other hand, the gray-body modification $\gamma_l^{\rm gray}$ follows similarly from \eqref{eq:rapprox} and reads
\begin{equation}
    \gamma_l^{\rm gray}=+\int\frac{{\rm d}\nu}{e^{\frac{\nu}{\theta}}-1}\Bigg|_{\xzero=0}\left[\frac{\pi \left(8 (\lk+3)-135 \nu^2\right)f_l^c(\nu)}{108 (\lk)^{3/2} \left(f_l^c(\nu)+e^{\frac{9 \pi  \nu^2}{4 \sqrt{\lk}}}\right)^2}\right].
\end{equation}
We see that the contribution from the gray-body factor enhances the overall
detection rate based on the value of $\xzero$, which is consistent with the reduced reflection for larger values.
The form of $\gamma_l^{\rm black}$ is independent of the specifics in the gray-body factor because it only depends on the temperature. Therefore, 
in the remainder of Sec.~\ref{sec:examples},
we will focus on the contribution $\gamma_l^{\rm gray}$, whose behavior is completely determined by the
reflection coefficient. 

\subsubsection{S-wave analysis}

To isolate the influence of $\xzero$, we now consider s-waves with $l=0$.
This waves are ignorant of the centrifugal barrier and experience the reduced potential
\begin{align}
    v_0(x)=\left(1-\frac{1}{x}\right)\left(\frac{2+\xzero}{2x^3}-\frac{3\xzero}{2x^4}\right).
\end{align}
In Fig.~\ref{fig:rh}, we show the form of the potential $V_0(r/r_g)$ [equivalently $v_0(x)$] for different values of $\xzero=r_0/r_g$. We recall that for $x<3$, a higher value of $\xzero $ leads to a lowered potential, while for $x>3$, a non-zero $\xzero $ enhances the potential (because $(v_0-v_0|_{\xzero =0})\propto(x-1)(x-3)$). 

To extract the reflection coefficient, and thus the gray-body contribution, we expand the potential around its maximum and apply the formerly introduced formalism.
The position of the maximum $x_{\rm max}$ is now given by\footnote{Due to the missing $l$-dependent term in the potential, the polynomial \eqref{eq:rootdet} to determine the roots is now only quadratic which yields a much simpler value for the maximum.}
\begin{equation}
    x_{\rm max}=\frac{4+8\xzero +\sqrt{16-26\xzero +19\xzero ^2}}{6+3\xzero }.
\end{equation}
In the limit of general relativity, that is $\xzero\to0$, we find $x_{\rm max}\to 4/3$,
which is consistent with Fig. \ref{fig:rh}. Then, the WKB frequency becomes
\begin{equation}
    \Omega_0(x;\nu)=\left(1-\frac{1}{x}\right)\left(\frac{2+\xzero }{2x^3}-\frac{3\xzero }{2x^4}\right)-\nu^2,
\end{equation}
and its roots coincide by definition with the turning points.
With this knowledge, we take \eqref{eq:kl} and derive $K_0(\nu)$ 
\begin{equation}\label{eq:fullK}
    K_0(\nu)=-i\frac{(x_{\rm max}-1)(2x_{\rm max}+x_{\rm max}-3\xzero)-2\nu^2x_{\rm max}^5}
   {\sqrt{-12 ({\xzero }+2) x_{\rm max}^5+40 (2 \xzero +1) x_{\rm max}^4-90 \xzero  x_{\rm max}^3}}.
\end{equation}
The form of $K_0(\nu)$ shows that the $\xzero$ dependence is not so easy to determine. Therefore, we need to particularize our analysis to retrieve more information about the role of $\xzero$. Since the astrophysical case encompasses to navigate within the small-$\xzero $ limit, we expand \eqref{eq:fullK} to first order in $\xzero $,
\begin{equation}\label{eq:k0exp}
    K_0(\nu)= i\frac{256\nu^2-27}{216\sqrt{2}}+i\frac{2816\nu^2-27}{3456\sqrt{2}}\xzero +\mathcal{O}(\xzero ^2).
\end{equation}
We see that the correction due to $\xzero$ increases the frequency $K_0(\nu)$ unless $\nu\lessapprox 0.098$.
When plugging \eqref{eq:k0exp} into \eqref{eq:reflexion}, the exponential factor also increases and, thus, causes a decrease in the reflection through $\xzero$, which implies an increase
of the gray-body factor. 
More precisely, to first order in $\xzero $ we find
\begin{equation}\label{eq:r0reflexion}
   |R_0(\nu)|^2= \frac{1}{1+f_0(\nu)}- \xzero \,\frac{\pi\,f_0(\nu)\left(2816 \nu^2-27\right)}{1728 \sqrt{2}
   \left(1+f_0(\nu)\right)^2}+\mathcal{O}(\xzero ^2),
\end{equation} 
with $f_0(\nu):=\exp\left(\tfrac{\pi (256\nu^2-27)}{108\sqrt{2}}\right)$. The negative sign in front of the correction term proves that the reflection is in general reduced unless $\nu\lessapprox 0.098$. Since our approximation breaks anyways down at small frequencies, we would not be able to trust the results for such small frequencies. 

\subsection{Almost-remnant configuration}\label{sec:arc}

This subsection covers the last stages before the temperature drops to zero and,
in principle, the remnant forms. This physical scenario corresponds to
the limit where $\xzero \nearrow 1$, that is, $r_g\searrow r_0$, and thus
the gravitational radius asymptotes to the position of the transition surface.
In such limit, the temperature of the horizon \eqref{eq:hawkingtemp} drops to zero. As such, we would be left with a stable remnant of radius $r_g=r_0$, which represents a configuration that cannot be realized in general relativity. For such a remnant, the exact potential reads
\begin{align}\label{eq:remnantpot}
    V_l(r)=\left(1-\frac{r_g}{r}\right)\left(\frac{l(l+1)}{r^2}+\frac{3r_g}{2r^3}\left(1-\frac{r_g}{r}\right)\right).
\end{align}
By definition, the remnant potential is independent of $r_0$ and, thus, becomes unsuitable to retrieve information about quantum-gravity corrections in the emitted spectrum. In fact, the remnant itself admits no Hawking effect because its temperature vanishes. Therefore, we focus on the moment close to the formation of the remnant,
such that the scale $\xzero$ remains in our analysis. That is,
we consider the potential for $\xzero\nearrow1$, which physically is translated to $r_g\searrow r_0$, and corresponds to the end stages of the evaporation. In this limit, the potential can be expanded as
\begin{equation}\label{eq:rempotexp}
    v_l(x)=\left(1-\frac1x\right)\left(\frac{(3+2\,\lk)x-3}{2x^4}-\frac{x-3}{2x^4}\epsilon\right)+\mathcal{O}(\epsilon^2),
\end{equation}
where we defined $\epsilon:=1-\xzero$ such that $\epsilon>0$.
This form allows us to study the approach to the remnant in a sensible manner. 

\subsubsection{Large $l$ analysis}

To understand the reflection coefficient for the almost-remnant state
for large angular mode number $l$, we repeat the previous steps in Sec.~\ref{sec.apbhll}.
That is, we expand the location of the maximum \eqref{eq:max}
for large $l$,
\begin{equation}
    x_{\rm max}=\frac32+\frac{1}{12\lk}-\frac{\epsilon}{4\lk}+\mathcal{O}\left(\frac{1}{\sqrt{(\lk)^3}}\right),
\end{equation}
which agrees exactly with the already found expression for the astrophysical black hole \eqref{eq:max1} when we replace $\epsilon=1-\xzero$. As a direct consequence, we find for $K_l(\nu)$
\begin{equation}
   K_l(\nu)= -i\frac{\sqrt{\lk}}{6}+i\frac{243\nu^2-8+8\epsilon}{216}\frac{1}{\sqrt{\lk}}+\mathcal{O}\left(\frac{1}{\sqrt{(\lk)^3}}\right),
\end{equation}
which obviously yields the $K_l(\nu)$ for the astrophysical black hole in \eqref{eq:expKLbig}. Therefore, by \eqref{eq:reflexion}, the reflection coefficient must be given by the same expression. Hence, the evaporation, at least for large $l$ seems to be consistent until the remnant is eventually formed. For the sake of completeness, let us derive $|R_l(\nu)|^2$ when expanded for small $\epsilon$
\begin{equation}
    |R_l(\nu)|^2=\frac{1}{h_l^o(\nu)+1}-\frac{\pi\,  \epsilon \,2 h_l^c(\nu)}{27 \sqrt{\lk} \left(h_l^c(\nu)+1\right)^2}+\mathcal{O}(\epsilon^2),
\end{equation}
where, similarly to the astrophysical black-hole scenario, we have defined the functions in the $\lambdabar$-independent
and in the $\lambdabar$-correction term as
\begin{eqnarray}
    h_l^o(\nu)&:=&\exp\left(\frac{\pi  \left(243 \nu^2-8\right)}{108 \sqrt{\lk}}-\frac{\pi  \sqrt{\lk}}{3}\right),\\
  h_l^c(\nu)&:=&  \exp \left(\frac{\pi  \left(81 \nu^2+4\right)}{324
   (\lk)^{3/2}}-\frac{\pi  \left(243 \nu^2-8\right)}{108 \sqrt{\lk}}+\frac{\pi  \sqrt{\lk}}{3}\right).
\end{eqnarray}
Here, we observe a difference in the correction term. Although formed through the same $K_l(\nu)$, the expansion for astrophysical black holes is only valid for small $\xzero$, whereas the remnant assumes large values. Again, the correction term becomes $\nu$-independent at $\mathcal{O}((\lk)^{-1/2})$, whilst featuring a negative sign. Thus, in this limit, we do not need a particular $\nu$-value to perceive the depletion in the reflectivity. However, the higher orders in $l$ become $\nu$-dependent and, for the minus sign to remain, we find the threshold value $\nu^2<\frac{8\lk}{135}$. Translating this back into using $\xzero$ instead of $\epsilon$, we confirm that a larger $\xzero$ leads to a {more} decreased reflection coefficient,
and thus to an increase of the gray-body factor.

\subsubsection{S-wave analysis}

For s-waves in the almost-remnant scenario, we find that the experienced potential simplifies dramatically. For the exact remnant configuration, $\Omega_0(x,\nu)$ reduces to the simple form
\begin{equation}
 \Omega_0(x;\nu)=\left(1-\frac{1}{x}\right)^2\left(\frac{3 }{2x^3}\right)-\nu^2,
\end{equation}
which, by definition, has no dependence on $\xzero$. As argued above, we are interested in the approach to the remnant case. Therefore, we turn to analyze the final stages of evaporation and consider the potential \eqref{eq:rempotexp} for $l=0$, 
\begin{equation}
    v_0(x)=\left(1-\frac1x\right)\frac{3x-3-\epsilon(x-3)}{2x^4}.
\end{equation}
At this point, we can proceed similarly to the astrophysical case before,
and determine the position of the maximum of this reduced potential $v_0(x)$,
\begin{equation}
    x_{\rm max}=\frac{12-8\epsilon+\sqrt{9-12\epsilon+19\epsilon^2}}{9-3\epsilon}.
\end{equation}
Up to this point, everything in the almost-remnant case is identical with the astrophysical black hole, since we only substituted our smallness parameter $\epsilon$
in the different expressions. Now, by taking the appropriate limit $\epsilon\to0$,
we can compute the function $K_0(\nu)$ evaluated at $x_{\rm max}$ to be 
\begin{equation}
    K_0(\nu)=i\frac{3125\nu^2-162}{675\sqrt{6}}-i\epsilon\frac{125\sqrt{2}\nu^2}{81\sqrt{3}}+\mathcal{O}(\epsilon^2).
\end{equation}
We learn that the $\lambdabar$-correction causes a depletion of $K_0(\nu)$. As before, we calculate its influence on the reflection coefficient to understand how the spectrum of the black hole behaves close to the final state of the remnant, which formally occurs at $\epsilon=0$. Using $K_0(\nu)$ in \eqref{eq:reflexion}, we find, for small $\epsilon$,
\begin{equation}
    |R_0(\nu)|^2=\frac{1}{h_0(\nu)+1}\left(1-\epsilon\frac{250\sqrt{2}\pi h_0(\nu)\nu^2}{81\sqrt{3}(h_0(\nu)+1)}\right)+\mathcal{O}(\epsilon^2),
\end{equation}
with $h_0(\nu):=\exp(\tfrac{\sqrt{2}\pi(3125\nu^2-162)}{675\sqrt{3}})$. We see that the larger $\xzero$, that is the smaller $\epsilon$, the larger becomes the reflection coefficient. This is evident from the potential analysis with restored units, because $r_g\searrow r_0$ during evaporation, which means that, effectively, the gravitational radius decreases, and so does the
almost-remnant potential \eqref{eq:remnantpot} when $l=0$. 

\section{The evaporation process}\label{sec.eva}

As derived in Section \ref{sec:tunneling} by making use of the tunneling approach, the temperature of the horizon is given by
relation \eqref{eq:hawkingtemp}, which we repeat here for convenience,
\begin{align}\label{eq.temperature}
    T= \frac{\hbar}{4\pi r_g}\sqrt{1-\frac{r_0}{r_g}}.
\end{align}
In order to understand the thermodynamics for the evaporation process,
we work in the parameter
space $(r_g,r_0)$, such that the temperature is a state function $T=T(r_g,r_0)$.

For the Schwarzschild black hole $(r_0\to 0)$, as a function of the radius of the horizon, the temperature
$T=T(r_g,0)$ is monotonically decreasing and diverges $T\to\infty$ as $r_g\to 0$. Therefore,
following the standard interpretation, given a black hole with a certain value $r_g$,
the black hole radiates energy away, i.e., loses mass. Logically, this implies a reduction of $r_g$, and thus leads to an increase of the
temperature. In the limit $r_g\to 0$, this process culminates into a complete evaporation of the black hole, achieving an infinite temperature, and
leaves the usual question about the information loss.

In contrast to the Schwarzschild case, and 
assuming a fixed value of $r_0>0$,
$T=T(r_g,r_0)$ is not a monotonic function of $r_g$. In fact, it has a finite value for all $r_g\geq r_0$:
It reaches a global maximum $T=\hbar/(6\sqrt{3}\pi r_0)$ for $r_g=3r_0/2$,
while it vanishes for $r_g\to r_0$ and for $r_g\to\infty$. Therefore, given an initial black hole with $r_g>r_0$, the radiated energy would shrink the size of the horizon, though in the limit $r_g\to r_0$, the radiation would
cease because the temperature of the horizon tends to zero. Therefore, the black hole would not completely
evaporate and, instead, it would leave behind a remnant with $r_g=r_0$ and vanishing temperature. 

However, since $r_0$ is defined as $r_0=\lambdabar r_g$,
instead of a constant $r_0$,
one could consider a constant $\lambdabar$ during the evaporation process.
Then, as $r_g$ shrinks, $r_0$ will also diminish. In such a case,
the evaporation of the horizon would tend, as in the Schwarzschild case,
to an object of vanishing gravitational radius and infinite temperature.
Therefore, the key issue is which of the two natural parameters,
$r_0$ or $\lambdabar$, is universal for all the solutions and
thus it is kept constant through the evaporation process.
However, we note that $\lambdabar=$constant simply provides a correction to the temperature
that can be absorbed in $\hbar$ (i.e., $\hbar\leftrightarrow\hbar\sqrt{1-\lambdabar}$).
Thus, in such scenario, one obtains exactly
the same qualitative evolution (and also entropy and time of evaporation)
as in GR, but with a rescaled value of $\hbar$. Therefore, in the following,
we will consider that $r_0$ is kept constant during the evaporation process,
and the results corresponding to the evaporation with a constant $\lambdabar$
are included in the GR limits below { with the corresponding rescaling
of $\hbar$. This choice is consistent with the existence of a positive minimum eigenvalue of the area operator in loop quantum gravity,
which is the theory that initially motivated these effective models \cite{Alonso-Bardaji:2021yls,Alonso-Bardaji:2022ear}.}

Let us therefore study the implications for the entropy of the horizon
and the evaporation time.
In order to construct the entropy, we need to define the energy (mass)
contained inside the horizon. A natural definition is given by the Hawking
energy\footnote{As shown in Ref.~\cite{Alonso-Bardaji:2022ear}, for this geometry the Hawking energy reads $\frac{1}{2r}[(r_g+r_0)r-r_0 r_g]$.} evaluated at the horizon, as it measures the amount of energy
enclosed by a given surface. Hence, we define
the energy contained inside the horizon as $E_{BH}:=r_g/2$.
Another key thermodynamic quantity is the pressure but, by definition, 
the present model is a vacuum solution and thus we will state that
its corresponding pressure is zero.

In this way, we can directly use the
definition of entropy $\delta S=\delta E_{BH}/T=\delta r_g/(2 T)$,
where $\delta$ 
stands for a variation on the parameter space,
and write,
\begin{align}\label{eq.entropy}
    S&=\frac{2\pi}{\hbar}\int \frac{r_g^{3/2}\delta r_g}{\sqrt{r_g-r_0}}%
    =\frac{\pi r_g^2}{\hbar}\left(1+\frac{3r_0}{2r_g}\right)\sqrt{1-\frac{r_0}{r_g}}+\frac{3\pi r_0^2}{4\hbar}\,\ln\left(\frac{\sqrt{r_g}+\sqrt{r_g-r_0}}{\sqrt{r_g}-\sqrt{r_g-r_0}}\right),
\end{align}
where the integration constant has been fixed so that vanishing entropy corresponds to
the remnant phase $r_g\to r_0$ with vanishing temperature.
We note that this is in fact the configuration with minimum entropy,
and the value of $S$ increases with $r_g$.

If we write the expression for the entropy in terms of the area
of the horizon $A=4\pi r_g^2$, and perform a series
expansion for large $A$,
\begin{align}\label{correctedS}
S= \frac{A}{4\hbar}+\frac{r_0}{2\hbar}\sqrt{\pi A}+\frac{3\pi}{8\hbar} r_0^2\ln(A)+{\cal O}(A^{0}),
\end{align}
we see that the leading-order correction to the Bekenstein-Hawking entropy
goes as the square root of the area, but it also acquires a logarithmic
term in the area, which appears in a wide variety of contexts.

Another interesting quantity we can study is the lapse of time a given black hole
takes to reach its corresponding equilibrium state. For such a purpose, we will make use of the Stefan-Boltzmann law in the form
\begin{equation}
\frac{\delta E_{BH}}{\delta\tau}=-\sigma A T^4,
\end{equation}
where, in the present units $\sigma:=\pi^2/(60\hbar^3)$
and $\tau$ is a formal time variable in the parameter space.
Note that this expression only takes into account the pure black-body spectrum and disregards the gray-body factor.
Therefore, in this approximated computation, we are overestimating the radiated power
and thus we will obtain a minimum bound for the evaporation time.

Since, as explained above, $\delta E_{BH}:=\delta r_g/2$,
one obtains the rate of contraction of the horizon,
\begin{align}\label{eq.evaporation}
    \frac{\delta r_g}{\delta\tau}=-\frac{\hbar}{1920\pi r_g^2}\left(1-\frac{r_0}{r_g}\right)^2.
\end{align}
\begin{figure}
    \centering
    \includegraphics[width=0.8\textwidth]{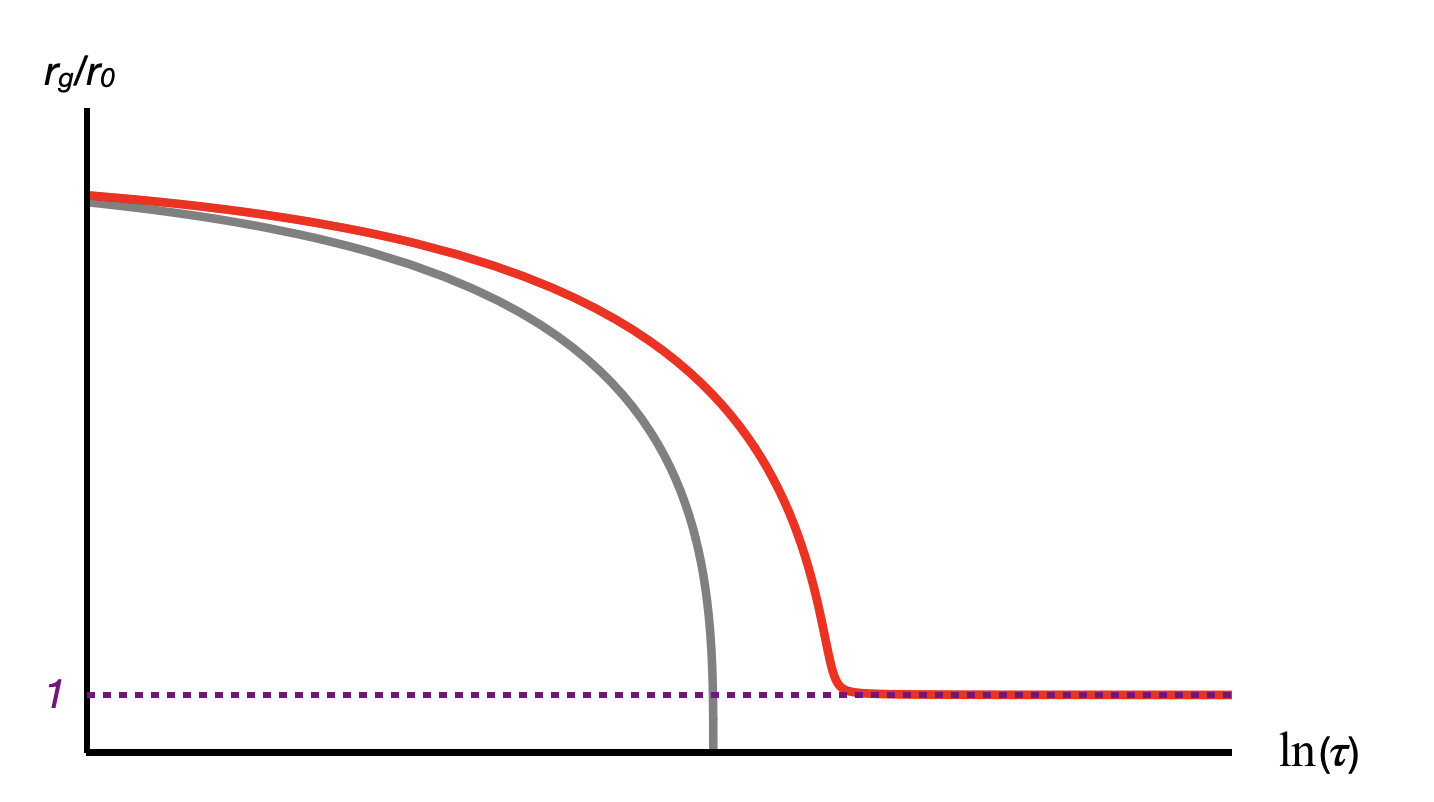}
    \caption{The decrease of the black-hole size during the evaporation process (red) and its classical trajectory (gray) as a function of $\ln(\tau)$. The behavior is qualitatively similar until very small masses. However, while the classical trajectory ends at a finite value of $\tau$, the red one tends asymptotically to the infimum $r_g=r_0$ (shown as a dashed purple line).}
    \label{fig.m(t)}
\end{figure}%
From this expression, it is straightforward to see that
the function $r_g(\tau)$ is monotonically decreasing. In addition, as shown in Fig.~\ref{fig.m(t)},
$r_g(\tau)$ has an inflection point at $r_g=2r_0$, where the evaporation starts to decelerate, and
it asymptotically approaches $r_g=r_0$ from above. 
In fact, we can analytically integrate the above expression
in the intervals $\tau\in[\tau_i,\tau_f]$ and $r_g\in[R_f,R_i]$, and obtain
\begin{align}\label{eq.evaporationtime}
    \frac{\hbar(\tau_f-\tau_i)}{7680\pi}=\frac{R_i^3-R_f^3}{3}-\frac{1}{4}\left(\frac{R_i^4}{R_i-r_0}-\frac{R_f^4}{R_f-r_0}\right)+\frac{r_0}{2}(R_i^2-R_f^2)+r_0^2(R_i-R_f)
    +r_0^3\ln\left(\frac{R_i-r_0}{R_f-r_0}\right),
\end{align}
with $\tau_i\leq \tau_f$ and $r_0<R_f\leq R_i$. 
As we stated above, this result is a lower bound for the evaporation time, but, since it is already infinite for any $r_0>0$, we do not need to include the gray-body factor. 

For large black holes $R_i\gg r_0$, we find that the difference with the ``classical'' time to achieve $R_f=2r_0$ is quadratic in the initial radius, i.e., ${\tau_f- \tau_f|_{r_0\to 0}} \approx 10^{5}r_0R_i^2/\hbar$. After that, the complete evaporation would occur in GR, whereas here the stable state of the remnant is never reached.

{The infinite evaporation time is in complete accordance with the third law of 
black-hole thermodynamics, which states that, starting from a finite temperature 
state, the state of zero temperature cannot be achieved in a finite amount of steps.
This fact is reflected by the infinite evaporation time because the actual remnant solution does correspond to a zero temperature configuration and marks mathematically the asymptotic endpoint of the evaporation process.}
 
{
In summary, when considering $r_0$ to be the fundamental constant of the theory, as motivated, for instance,
by the infimum of the area operator in loop quantum gravity, the effective
model predicts the formation of stable black-hole remnants. These solutions, however, seem to be a set of
measure zero that cannot be attained dynamically. This fact resonates with the conclusions in \cite{Borges:2023fub},
where the attainability of the remnant solution has been studied from the perspective of extremal horizons.}

\section{Conclusions}\label{sec:conclusions}

We have analyzed the radiative and thermodynamic properties of the nonsingular black-hole geometry \eqref{eq.metric} presented
in Refs.~\cite{Alonso-Bardaji:2021yls,Alonso-Bardaji:2022ear}. This geometry is a solution of the equations of motion generated by a Hamiltonian constraint, which is deformed with
respect to the Hamiltonian of general relativity, in the sense that it contains certain
functions, parametrized by a dimensionless parameter $\lambda$, which can be understood
as encoding loop-quantum-gravity corrections. However, the model is more general and it is
not derived in the context of loop quantum gravity, so it could also model any other type
of process that leads to a resolution of the classical singularity.

The geometry is qualitatively similar to that of the Schwarzschild black hole, in the sense
that it is asymptotically flat and it presents a Killing horizon at $r=r_g$. However, instead
of the singularity, this geometry presents a surface at $r=r_0$ that links the black-hole region
to a white-hole region (see Fig.~\ref{fig.diagram}). The constant $r_0$ is directly related to the parameter $\lambda$,
and the Schwarzschild solution corresponds to $r_0\to0$.

In this paper we have explicitly derived the radiative spectrum of this regular
black-hole horizon.
As for any other body, the radiance of the horizon can be separated into two parts:
the black-body part, which is completely characterized in terms of the temperature,
and the gray-body factor. This factor measures the departure of the spectrum from
that corresponding to a perfect black body, and it can be understood as originating
by those modes that are emitted by the horizon, but bounce at the potential barrier,
and are thus absorbed back by the horizon.

We computed the temperature of the geometry \eqref{eq.metric}
by making use of the tunneling method.
As can be seen in \eqref{eq:hawkingtemp}, the factor $r_0$ reduces the temperature of the horizon
as compared to the Schwarzschild case (for black holes of equal size $r_g$). Therefore, we have a colder body, that emits
less radiation.
Besides, the height of the potential barrier is also diminished by $r_0$ (see Fig.~\ref{fig:rh}), which implies
a lower value of the gray-body factor as compared to Schwarzschild.
We have explicitly computed the gray-body factor by making use of the standard WKB
approach that provides the gray-body factor in terms of the reflections coefficient.
We have analyzed in detail its specific form for several cases of interest. More precisely,
we have studied large astrophysical black holes (with $r_0\ll r_g$) as well as
almost-remnant configurations (with $r_0\approx r_g$). In particular, we considered
large-$l$ modes and s-waves $(l=0)$, and derived the specific properties of the gray-body
factor.
All in all, we can summarize our results as follows: compared to a Schwarzschild black hole, the presence of the scale $r_0$ results in a colder but less gray spectrum. 

The absence of a singularity requires
the violation of the singularity theorems \cite{carballo2020opening}, which is here achieved
by a turnaround of the future trapping into an anti-trapping.
Due to the spherical symmetry, this class of singularity-free spacetimes features a smaller
future trapped region that ventures into a past trapped region.
From the perspective of an observer
outside the horizon, this results in a lower temperature and surface gravity. It seems therefore
that, through the attenuation of the curvature singularity, the overall gravitational
pull is also reduced, which culminates in a reduced surface gravity as well as effective potential.
Given equal sizes, this suggests the generic feature that nonsingular
black holes admit a purer black-body spectrum than their singular
siblings predicted by general relativity.

Finally, we have studied the thermodynamic properties of this model.
In order to find the entropy of the horizon, one needs to define the energy
contained within it. A natural magnitude is the Hawking energy evaluated
at the horizon. With such definition, and performing an expansion for large
areas, we find corrections to the Bekenstein-Hawking
entropy that go as the square root and the logarithm of the horizon area
(see Eq.~\eqref{correctedS}). In addition, we conclude that,
if one assumes that the parameter $r_0$ is kept constant
as the horizon radiates, the end state of the evaporation
process corresponds to a stable remnant with vanishing temperature and entropy. 

{ This remnant configuration is reached in the vanishing temperature case only as an asymptotic endpoint after an infinite time-span. This result confirms the validity of the third law of black-hole thermodynamics for our model. Nevertheless, the presence of the remnant as a solution of the model remains important for further research. This remnant may encode enough information to address the information-loss paradox, but only a careful study of the backreaction will unravel its dynamical attainability.}

\begin{acknowledgments}
This work has been supported by the Basque Government Grant
\mbox{IT1628-22} and by the Grant PID2021-123226NB-I00 (funded by
MCIN/AEI/10.13039/501100011033 and by ``ERDF A way of making Europe'').
\end{acknowledgments}

\appendix

\section{Other comparisons}
\label{app.others}
In the main text, when commenting the effects of $r_0>0$ with respect to its corresponding GR counterpart  ($r_0\to 0$), we have assumed black holes of the same same `size', that is, with the same gravitational radius $r_g$. However, the additional constant $r_0$ of the effective theory permeates the whole spacetime, and some other comparisons might be relevant. For instance, one could consider black holes with the same asymptotic (ADM) mass
or with the same temperature (surface gravity). Indeed, since the Komar and Hawking masses do not coincide in this effective theory \cite{Alonso-Bardaji:2022ear} (recall that Einstein's equations are not satisfied), these two cases deliver different corrections. In contrast to what we analyzed in the main text (Sec.~\ref{sub:msd}), the maximum of the effective potential (and thus the critical frequency and the reflection coefficient) increase with $r_0$
for both these cases. For simplicity, let us work again with $\lambdabar=r_0/r_g$.

\begin{figure}[h]
    \centering
        \hfill
        \includegraphics[width=.49\textwidth]{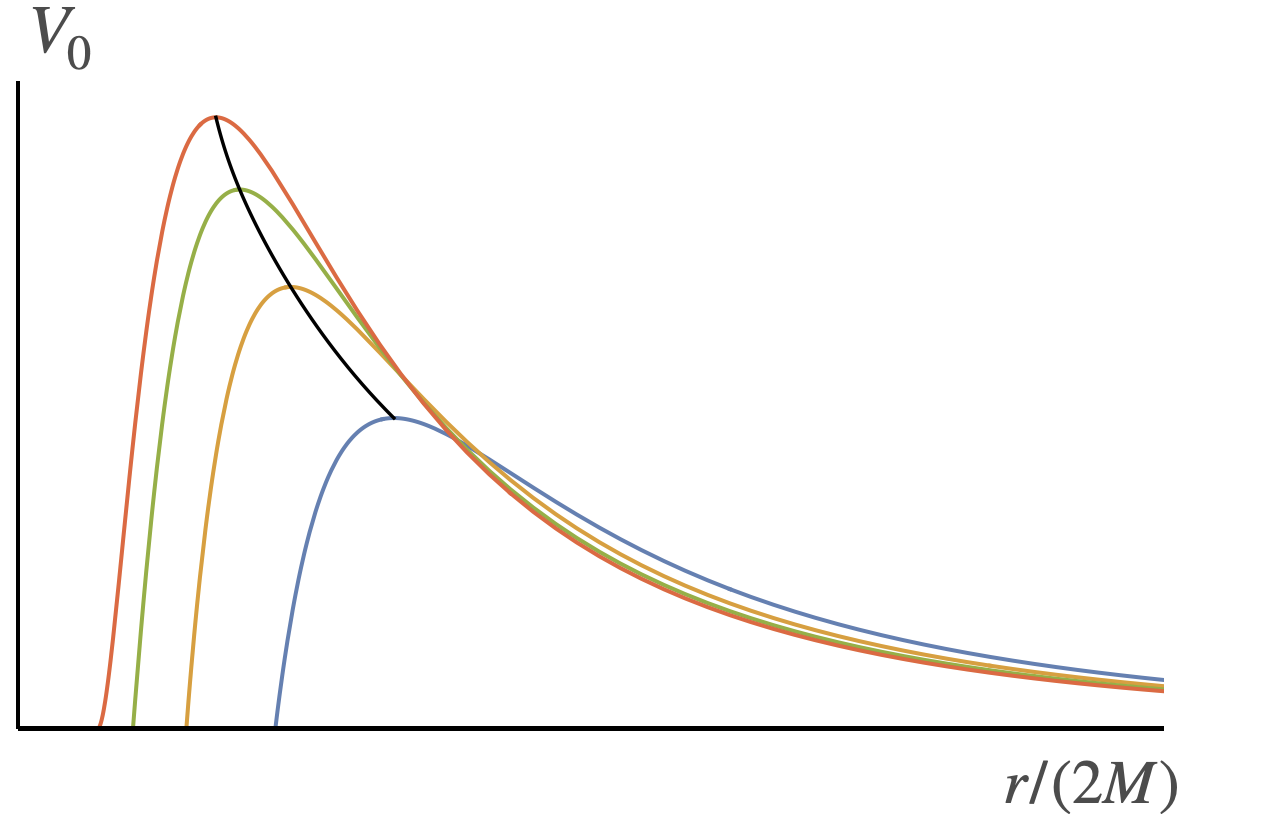}
        \hfill
        \includegraphics[width=.49\textwidth]{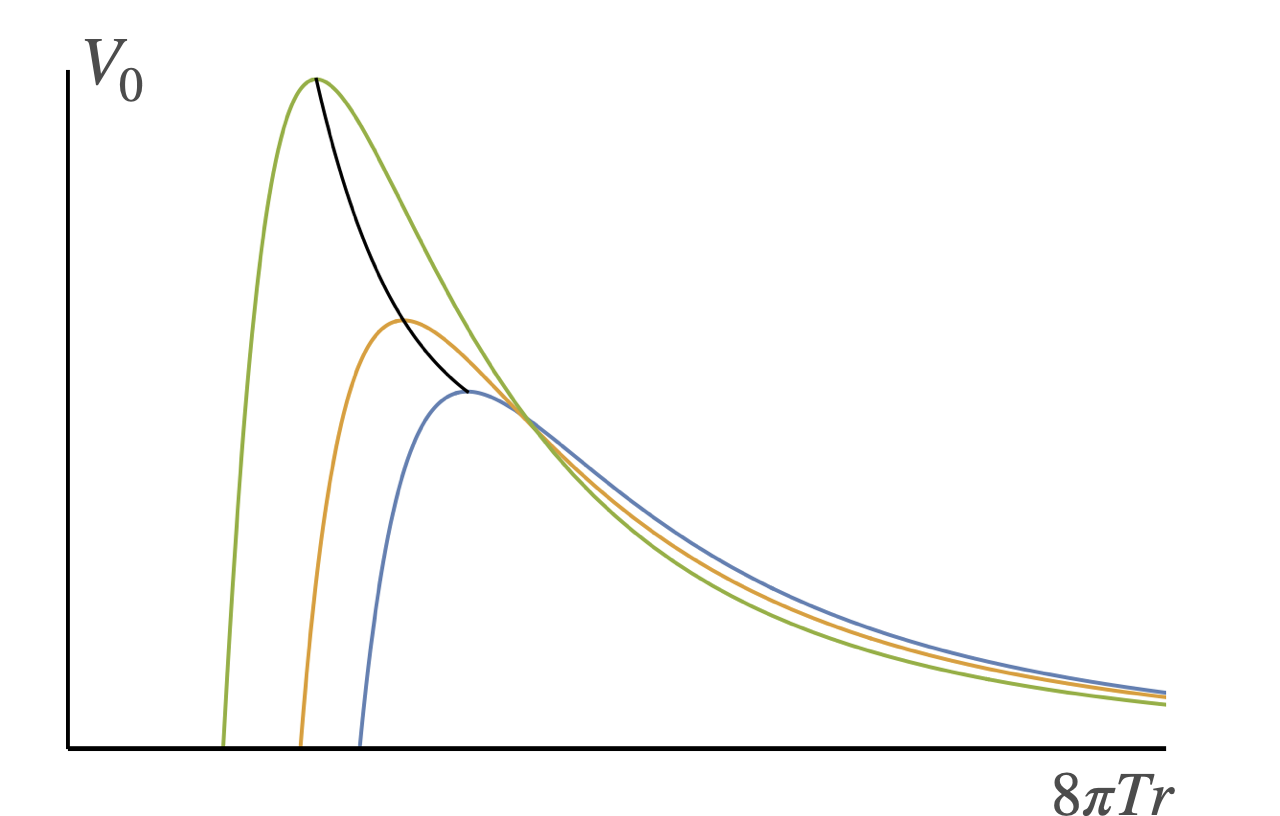}
        \hfill
     \caption{Left: Regge-Wheeler potential with $l=0$ for black holes of same ADM mass ($M=1/2$) with $\lambdabar\in\{0,1/3,2/3,1\}$ in $\{$blue, orange, green, red$\}$, respectively. The black line
     shows the position of the maximum for all the possible values of $\xzero$. 
     Right: Regge-Wheeler potential with $l=0$ for black holes of the same temperature ($T=\hbar/(8\pi)$) with $\lambdabar\in\{0,1/3,2/3\}$ in $\{$blue, orange, green$\}$, respectively. }
     \label{fig:ADM&T}
\end{figure}

In terms of the ADM mass $M:=(1+\lambdabar)r_g/2$ \cite{Alonso-Bardaji:2022ear},
the potential \eqref{eq:aform} reads
\begin{align}
    V_l(r,M)=\left(1-\frac{2M}{(1+\lambdabar)r}\right)\left(\frac{l(l+1)}{r^2}+\frac{2+\lambdabar}{1+\lambdabar}\frac{M}{r^3}-\frac{6M^2\lambdabar}{(1+\lambdabar)^2r^4}\right).
\end{align}
For a given $M$, $r_g$ shrinks as we increase $\xzero$.
This implies that, as $\xzero$ increases, the height of the maximum of the potential increases,
and its position is found at smaller radii (see Fig.~\ref{fig:ADM&T}).

Let us turn now to the equal temperature case.
In terms of the temperature \eqref{eq.temperature}, the potential \eqref{eq:aform} reads
\begin{align}
    V_l(r,T)=\left(1-\frac{\pi\hbar\sqrt{1-\lambdabar}}{T r}\right)
    \left(\frac{l(l+1)}{r^2}+(2+\lambdabar)\pi\hbar\frac{\sqrt{1-\lambdabar}}{2T r^3}-\frac{3\pi^2\hbar^2\lambdabar (1-\lambdabar)}{2T^2 r^4}\right).
\end{align}
The qualitative behavior is the same as in the constant ADM mass case above.
As can be seen in Fig.~\ref{fig:ADM&T}, the smaller $\xzero$,
the shallower the potential.

\section{Location of the maximum of the potential}\label{app:rmax}

To determine the location of the maximum of the potential \eqref{eq:aform}, we have to solve a cubic polynomial equation of the form
\begin{equation}\label{eq:rootdet}
4 \lk r^3+ 3 ( 4 r_g + r_0-2 \lk) r^2 - 4 (r_0 + r_g (4 + 3 r_0)) r+15 r_g\, r_0=0.
\end{equation}
For these polynomials there exists a general way to solve them, which yields three roots: one real and two potentially complex. For the above case, we find exactly one real root, that is, 
\begin{equation}\label{eq:max}
    r_{\rm max}=\frac{1}{12\lk}\left(6\lk-12r_g-3r_0-\alpha-\frac{\Delta_0}{\alpha}\right),
\end{equation}
where we have defined 
\begin{equation}
\alpha:=\sqrt[3]{\frac{\Delta_1+\sqrt{\Delta_1^2-4\Delta_0^3}}{2}},
\end{equation}
as well as the resultants $\Delta_0$ and $\Delta_1$ to be
\begin{eqnarray}
    \Delta_0&=&9 (-2 l (l+1) + 4 r_g + r_0)^2 + 48 l (l+1) (r_0 + r_g (4 + 3 r_0)),\\
    \Delta_1&=&18 (360 (l (l + 1))^2 r_g \,r_0 + (-2 l (l+1) + 4 r_g + r_0)^2 \\
    &&\hspace{1cm}+ 
   24 l (l+1) (-2 l (l+1) + 4 r_g + r_0) (r_0 + r_g (4 + 3 r_0)))\nonumber.
\end{eqnarray}
The other two roots can be found by the rescaling $\alpha\to z^s\alpha$,
with $z$ being a primitive cube root of the unity $z=\frac{-1+i\sqrt{3}}{2}$, while $s\in\{1,2\}$.

\bibliography{biblio}

\end{document}